\documentclass[aoas,nameyear,dvips]{arximspdf}
\usepackage{dcolumn,multirow}
\usepackage{graphicx}

\doi{10.1214/09-AOAS255}
\volume{3}
\issue{4}
\pubyear{2009}
\firstpage{1403}
\lastpage{1447}

\makeatletter
\newcolumntype{d}[1]{D{.}{.}{#1}}

\makeatother

\begin{document}
\begin{frontmatter}

\title{Workload forecasting for a call center: Methodology and a case study}
\runtitle{Workload forecasting for a call center}

\begin{aug}
\author[A]{\fnms{Sivan} \snm{Aldor-Noiman}\corref{}\thanksref{t1}\ead[label=e1]{sivana@wharton.upenn.edu}},
\author[B]{\fnms{Paul D.} \snm{Feigin}\thanksref{t2}\ead[label=e2]{paulf@ie.technion.ac.il}}
\and\break
\author[B]{\fnms{Avishai} \snm{Mandelbaum}\thanksref{t2}\ead[label=e3]{avim@tx.technion.ac.il}}
\thankstext{t1}{Currently a Ph.D. candidate in Department of
Statistics, The Wharton School, University of Pennsylvania.}
\thankstext{t2}{Supported in part by The Israel Science Foundation
Grant 1046/04.}

\runauthor{S. Aldor-Noiman, P. D. Feigin and A. Mandelbaum}
\affiliation{University of Pennsylvania and Technion---Israel
Institute of Technology}
\address[A]{S. Aldor-Noiman\\
Department of Statistics\\
The Wharton School\\
University of Pennsylvania\\
Philadelphia, Pennsylvania 19104\\
USA\\
\printead{e1}} 
\address[B]{P. D. Feigin\\
A. Mandelbaum\\
William Davidson Faculty of\\
\quad Industrial Engineering
and Management\\
Technion--Israel Institute of Technology\\
Technion City, Haifa 32000\\
Israel\\
\printead{e2}\\
\phantom{E-mail:\ }\printead*{e3}}
\end{aug}

\received{\smonth{11} \syear{2008}}
\revised{\smonth{2} \syear{2009}}

%
\begin{abstract}
Today's call center managers face multiple operational
decision-making tasks. One of the most common is determining the
weekly staffing levels to ensure customer satisfaction and meeting
their needs while minimizing service costs. An initial step for
producing the weekly schedule is forecasting the future system loads
which involves predicting both arrival counts and average service
times.

We introduce an arrival count model which is based on a \textit{mixed}
Poisson process approach. The model is applied to data from an
Israeli Telecom company call center. In our model, we also consider
the effect of events such as billing on the arrival process and we
demonstrate how to incorporate them as exogenous variables in the
model.

After obtaining the forecasted system load, in large call centers, a
manager can choose to apply the QED (Quality-Efficiency Driven)
regime's ``square-root staffing'' rule in order to balance the
offered-load per server with the quality of service. Implementing
this staffing rule requires that the forecasted values of the
arrival counts and average service times maintain certain levels of
precision. We develop different goodness of fit criteria that help
determine our model's practical performance under the QED regime.
These show that during most hours of the day the model can reach
desired precision levels.
\end{abstract}

%
\begin{keyword}
\kwd{Call centers}
\kwd{QED regime}
\kwd[, ``]{square-root staffing}
\kwd[,'' ]{forecasting arrival count}
\kwd{exogenous variables}.
\end{keyword}
\end{frontmatter}

\section{Introduction}\label{sec1}
\setcounter{footnote}{2}

Many companies invest significant resources in order to provide high
quality customer service, with much or all customer interactions
based on telephone or internet access. Telephone Call Centers, and
their multimedia extensions called Contact Centers, support these
interactions between companies and their customers. Such service
centers accumulate vast amounts of data, which can be analyzed and
utilized for short-term operational decisions, medium-term tactical
decisions or long-term strategic decisions.

Over recent years, the service industry has expanded dramatically.
Estimates from 2005 indicated that call center costs exceeded
$\$300$ billion worldwide [Gilson and Khandelwal (\citeyear{mckinsey})]. It is estimated that
there are 4 million call center agents in the USA, 800 thousand
in the UK, 500 thousand in Canada and 500 thousand in India
[Holman, Batt and Holtrewe (\citeyear{cornell})].

One of the main challenges in operating a telephone call center is
determining staffing levels that meet future demand, given desired
levels of service quality and efficiency. A prerequisite for such a
task is the forecasting of the system workload over the periods of
the day, for several days in advance. The workload, or what is
technically called the \textit{offered-load}, depends on the arrival
process and the service times that each arrival (customer) requires.
For planning a staffing schedule, call center operators utilize
forecasts of the arrivals and service times at a sufficient
resolution---say for every half hour. Given that information,
jointly with some understanding of customer patience characteristics
[Zeltyn (\citeyear{sergeyPhD})] and increasingly prevalent software tools (e.g.,
4CallCenters Software,
\href{http://ie.technion.ac.il/serveng/4CallCenters/Downloads.htm}{http://ie.technion.ac.il/serveng/}
\href{http://ie.technion.ac.il/serveng/4CallCenters/Downloads.htm}{4CallCenters/Downloads.htm}),
an operator can determine the required number of agents for each period.

A common approach for achieving proscribed service quality and
efficiency [Zeltyn (\citeyear{sergeyPhD})] is via the \textit{square-root staffing
rule}. Assume for simplicity a constant arrival rate of $\lambda$
calls per minute, and let the average service time be $E[S]$ (in
minutes). The \textit{offered-load} $R$ is defined to be $R = \lambda
\times E[S]$: it is the average amount of work (in service time)
that arrives to the system (per unit of time). With a forecast of
$R$, one sets the number of agents to be $N=R+\beta\sqrt{R}$, for
some parameter $\beta$ (typically $-1\leq\beta\leq2$) that
reflects the required balance between service quality (short waiting
times, few abandonments) and service efficiency (utilization of
agents). Selecting a value for $\beta$ is the manager's way of
achieving a required call center performance (see
Section~\ref{section:beta}).

The purpose of the present paper is to describe an implementation of
the above described program, from statistical modeling of the
arrival and service processes, through using these models for
forecasting, and finally applying the forecasts to predict workload
and thus staffing requirements.

We use Gaussian linear mixed model formulations to describe both a
suitably transformed version of the arrival process and the average
service times. Mixed models allow us the much needed flexibility to
describe different seasonality effects using correlation structures
in both models. Moreover, they provide more realistic prediction
intervals than those obtained ignoring correlations in the series.
Mixed models also allow us to incorporate exogenous variables in a
``natural'' manner.

Since forecasting\footnote{In this paper we shall use the terms
forecasting and predicting interchangeably.} is an error prone
activity, we also analyze the ramifications of forecasting errors
on system performance, when compared with the desired level of
performance as determined under the QED regime.

This paper describes both the analysis of a particular call center and
the methodological approach that can be used as
a blueprint for other call centers. Significantly, the forecasting
models considered can be implemented using \textit{standard}
statistical software---such as SAS\tsup{\textregistered}/STAT
[SAS (\citeyear{SAS}) used here] or R [R Development Core Team (\citeyear{Rstat})].

We start with a review of past and recent studies that have been
conducted on call center arrival processes. We then describe our
case study data set (Section~\ref{section:data description}), and
explain the component models used for workload forecasting
(Sections~\ref{section:arrival process} and~\ref{section: service times}). Finally, we consider the forecasting of
offered-load, its application to staffing, and the effects of
forecasting error on performance, all in
Section~\ref{section:staffing}.

\section{Literature review}

For a review of research on the operational aspects of call centers,
readers are referred to Gans Koole and Mandelbaum (\citeyear{CCtutorial}). A more recent survey
focusing on the multiple disciplines required to support call center
research is Aksin, Armony and Mehrota (\citeyear{MorReview}).

\textit{Queueing models}: The operational reality of a
basic call center can be well captured by the $M_t/G/N + G$ queue
[Whitt (\citeyear{queueing})]. Here one assumes i.i.d. (independent and
identically distributed) customers and i.i.d. servers (agents), with
a time-varying arrival-rate of calls. Formally: (i) $M_t$
indicates that the arrival process is nonhomogeneous Poisson with
a deterministic arrival rate function $\lambda(t)$; (ii) The first
$G$ indicates that the service times are i.i.d., independent of the
arrival process, each distributed as a random variable (to be
generically denoted $S$) with cumulative distribution function
$G(x)$ and finite mean $\frac{1}{\mu} \equiv E(S)$;
(iii) $N$ is the number of servers, which is allowed to be a
time-dependent deterministic function; and (iv)~the second $G$
indicates that customers can abandon while waiting to be served;
customers' impatience times (times to abandon) are i.i.d.,
independent of the arrival process and the service times, and with
finite mean $\frac{1}{\theta}$.

The $M_t/G/N + G$ model is intractable analytically. Fortunately,
for practical purposes it is often reducible to the tractable
$M_t/M/N + M$ model [Brown et~al. (\citeyear{Statistical-Analysis})], in which
service time distribution is $\operatorname{exponential}(\mu)$ and patience is
$\operatorname{exponential}(\theta)$. In Section~\ref{section:staffing} we
describe how to use the $M_t/M/N + M$ model, often referred to as
Erlang-A, to assess the accuracy of our proposed forecasting method.

\textit{Call arrivals}: Early forecasting studies applied
classical Box and Jenkins, Auto-Regressive-Moving-Average (ARMA)
models, for example, the Fedex study [Xu (\citeyear{Fedex})] and L.~L. Bean
[Andrews and Cunningham (\citeyear{LLBeans})]. The latter study considers two arrival
processes, each with its own characteristics: it incorporates
exogenous variables along-side MA (Moving Average) and AR
(Auto-Regressive) variables, using transfer functions to help
predict outliers such as holidays and special sales promotion
periods. Antipov and Meade (\citeyear{Meade-Antipov}) also tackled the problem of
including advertising response and special calender effects by
adding exogenous variables in a multiplicative manner.

More recently, Taylor (\citeyear{Taylor}) applied several time series models to
two sources of data, including seasonal ARMA models, double
exponential smoothing methods for seasonality and dynamic harmonic
regression. His results indicated that, for practical forecasting
horizons (longer than one day), a very basic averaging model
outperforms all of the suggested more complex alternatives.

\textit{Beyond Poisson}: Recent empirical work has revealed
several important characteristics that underly the arrival process
of telephone calls:
\begin{itemize}
\item\textit{Time-variability}: Arrival rates vary temporally over
the course of a day, see Tanir and Booth (\citeyear{TanirBooth}). For example, peak hour
arrival rate can be significantly higher than the level of the
average daily arrival rate [Brown et~al. (\citeyear{Statistical-Analysis})];
\item\textit{Over-dispersion}: Arrival counts exhibit variance that
substantially dominates the mean value. This goes against the
assumption that the arrival process is Poisson. A mechanism that
accounts for over-dispersion was suggested by
Jongbloed and Koole (\citeyear{Koole-Jongbloed}). They proposed the Poisson mixture model
which incorporates a \textit{stochastic} arrival rate process to
generate the additional variability;
\item\textit{Inter-day correlation}: There is significant dependency
between arrival counts on successive days.
Brown et~al. (\citeyear{Statistical-Analysis}) suggest an arrival forecasting model
which incorporates a random daily variable that has an
autoregressive structure to explain the inter-day correlations;
\item\textit{Intra-day correlation}: Successive periods within the
same day exhibit strong correlations. This intra-day correlation was
modeled and analyzed by Avramidis, Deslauriers and L'Ecuyer (2004),
using Dirichlet distributions.
\end{itemize}

\textit{Forecasting methods}:
In recent years, technological advances have enabled researchers
to employ sophisticated Bayesian techniques to call forecasting.
These yield the forecasted arrival \textit{counts}, as well as their
\textit{distribution}, thus providing far more information than just
point estimates.

For example, Soyer and Tarimcilar (\citeyear{soyer-tarimcilar}) analyzed the effect of
marketing strategies on call arrivals. Their Bayesian analysis is
based on the Poisson distribution of arrivals over time periods
measured in days, with cumulative rate function. They model the rate
function using a mixed model approach. Their conclusion is that the
random effects model fits much better than the fixed effects model;
indeed, the data cannot be adequately described by assuming a fixed
model without some additional random variability source. The mixture
model also provides within advertisement correlations over different
time periods.

Another model of incoming call arrivals to a US Bank call center,
also employing Bayesian techniques, was proposed in
Weinberg, Brown and Stroud (\citeyear{JBayesian}). In that paper the authors use the
Normal-Poisson stabilization transformation to transform the Poisson
arrival counts into normal variates. The normally transformed
observations allowed them the necessary flexibility to incorporate
conjugate multivariate normal priors with a wide variety of
covariance structures. The authors provide a detailed description of
both the one-day-ahead forecast and within-day learning algorithms.
Both algorithms use Gibbs sampling techniques and
Metropolis--Hastings steps to sample from the forecast distributions.
These techniques, although very modern, nevertheless still require
long processing times (since the procedure requires meeting the
convergence criteria as well as ``overcoming'' the
auto-correlations between successive samples). For our proposed
model the run time is actually quite short since we implement it
using a two-stage approach (which will be discussed in detail in
Section~\ref{subsubsection:two stage model}).

The Bayesian model also requires carrying out sensitivity analyses
with respect to the hyper-parameters. This fine tuning of the
parameters can be a much more tedious and burdensome process than
the adjustments we implement in our model (described in
Section~\ref{section:arrival process}).

Taking a non-Bayesian approach, Shen and Huang (\citeyear{SVD}) analyzed and modeled
call center arrival data using a Singular Value Decomposition (SVD)
method. The SVD algorithm allowed them to visually analyze the data.
Expanding on the same idea, Shen et al. take advantage of
this technique to reduce the dimension of the data and also create a
prediction model which provides inter-day forecasting and an
intra-day updating mechanism for the arrival rate profiles
[Shen and Huang (\citeyear{SVDpredict})]. In a more recent paper
Shen and Huang (\citeyear{PossionSVDpredict}) incorporate the previous idea (i.e.,
SVD) but make use of Poisson regression to give better predictions
of the arrival counts and their prediction intervals.

In our model we demonstrate how to incorporate covariates. This
process of adding exogenous variables is quite ``natural'' and easy
under the mixed model settings. It is not quite clear how one would
add these exogenous variables under the Bayesian setting proposed by
Weinberg et al. or using the SVD approach by Shen and Huang.

Our arrival count prediction model is a natural extension of the
model which was initially explored by Brown et~al. (\citeyear{Statistical-Analysis}).
However, we do offer some key modifications, such as adding exogenous
variables and modeling both intra- and inter-day correlation using
an auto-regressive process.

For a detailed bibliography on the subject of forecasting telephone
call arrivals, as well as other call centers' related papers, readers
are referred to Mandelbuam (\citeyear{CCbib}).

\section{Case study data}\label{section:data description}
Our data originate from an ongoing research\break project that is
conducted at the Technion's SEE Laboratory SSE
(\href{http://ie.technion.ac.il/Labs/Serveng/}{http://}
 \href{http://ie.technion.ac.il/Labs/Serveng/}{ie.technion.ac.il/Labs/ Serveng/}). 
This project [Feigin et~al. (\citeyear{DataMocca})], entitled DataMOCCA (Data \mbox{MOdels} for Call
Center Analysis), has created a repository of multiyear histories
from several call centers, at the individual-call level. The present
study focuses on data from an Israeli Telecom company, as will be
now described. See the supplemental material for further details
about the data Aldor-Noiman, Feigin and Mandelbaum (\citeyear{supp}).

\subsection{Arrival process}

The call center handles calls from several main queues:
Private clients; Business clients; Technical Support problems; Foreign
languages queues; and a few minor queues. In general, queues are
operated by separate service provider groups. Almost $30\%$ of the
incoming calls join the Private customers queue, which is catered by
a dedicated team of 150 telephone agents. The load generated from
each of the remaining queues is much smaller (e.g., the
second largest queue is the Business queue, which generates 18$\%$
of the overall incoming load). Hence, we shall limit our analysis to
the Private queue (and bear in mind that our modeling techniques are
applicable to the other queues as well).

The Private queue's call center operates six days a week, closing
only on Saturdays and Jewish holidays. On regular weekdays,
operating hours are between 7~a.m. and 11~p.m. and on Fridays it closes
earlier, at around 4~p.m.

The data includes arrivals between mid-February, 2004 and December
31, 2004. We divide each day into half-hour intervals. There are two
alternative justifications for choosing a half-hour analysis
resolution: (a) currently, shift scheduling is carried out at this
resolution; and (b) from a computational complexity point of view,
taking shorter intervals significantly increases the computing time
for many models and may make their implementation completely
impractical.

In the sequel we consider, for each day, the 24 half-hourly
intervals between 10~a.m. and 10~p.m., as this is when the call center is
most active.

Note that if the arrival rate was \textit{very} inhomogeneous during a
particular half-hour interval, then using the average arrival rate
could lead to under-staffing. Specifically, the staff level assigned
to meet the average load would not be able to cope with the peak
load in that particular half-hour interval. We do basically assume
in the sequel that the within interval inhomogeneity is mild---it
is further evaluated in Section~\ref{section:halfhour}.

\begin{figure}

\includegraphics{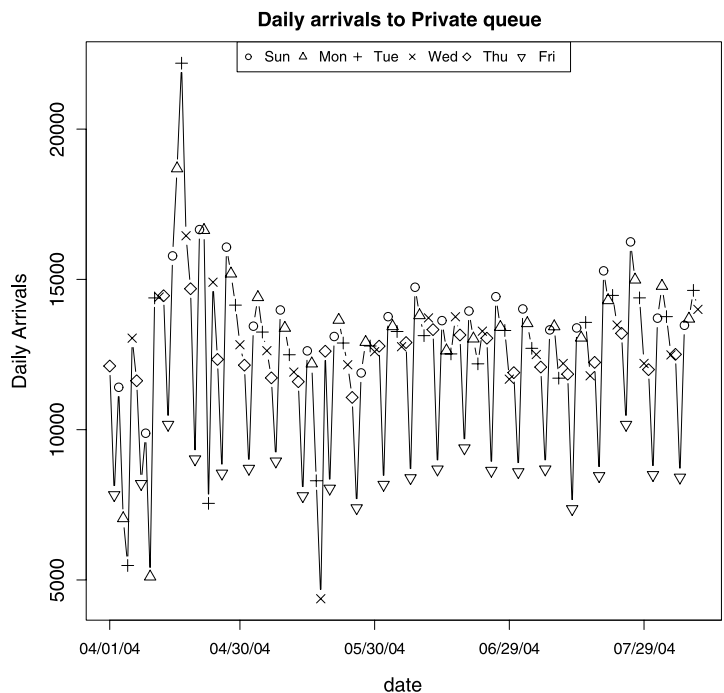}

\caption
{Daily arrivals, including holidays, to the Private queue between April
1st, 2004 and September 1st,
2004.}\label{fig:dailyTS}
\end{figure}

Figure~\ref{fig:dailyTS} demonstrates the arrival counts between
April, 2004 and September, 2004. Higher resolution analysis of this
graph gives rise to the following weekly patterns: Sundays and
Mondays have the highest arrival counts; the number of arrivals
gradually decreases over the week until it reaches its lowest point
on Fridays; and, there are quite a few outliers which occur in
April.

Examination of outlying observations singles out twenty-two days
with unusual arrival counts. Among these days, 17 days were holidays
and five days exhibited different daily patterns and unusual daily
volumes when compared to similar regular weekdays.

As mentioned earlier, April 2004 has an unusual weekly pattern. Out
of the list of twenty-two outliers, nine occur in April, which
explains the peculiar pattern that we see in Figure~\ref{fig:dailyTS}. These nine outliers can be attributed to Holidays
and a countrywide change in phone number prefixes. In conclusion,
the outlying days were excluded from the learning stage of our model
but were kept for later evaluation purposes.

%

\begin{figure}

\includegraphics{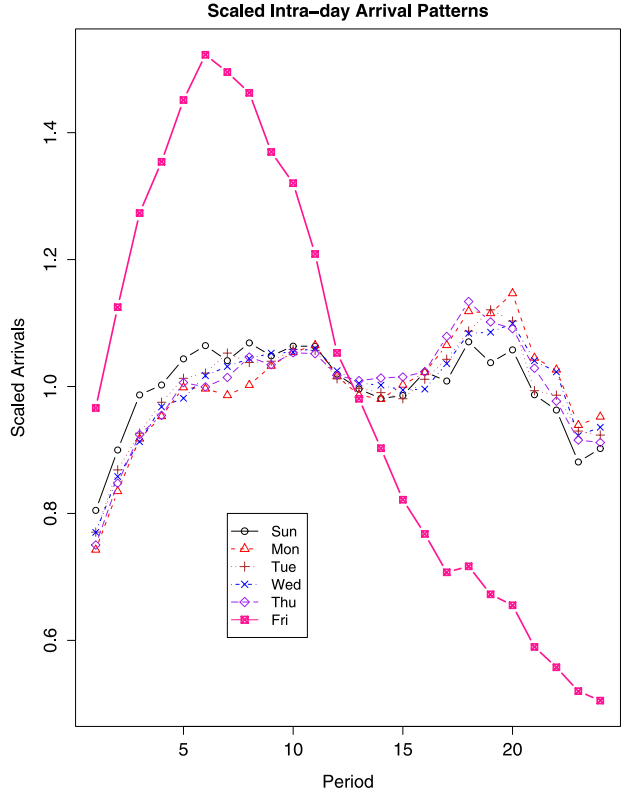}

\caption
{Normalized intra-day arrival patterns---averaged over weeks and smoothed.}\label{fig:Scaled Intra-day arrivals}
\end{figure}

Study of intra-day arrival patterns for regular days also reveals
some interesting characteristics:
\begin{itemize}
\item The weekdays Monday through Thursday have a similar
pattern. Figure \ref{fig:Scaled Intra-day arrivals}
illustrates the last fact by depicting the \textit{normalized} weekday
patterns---each half hour is divided by the mean half-hour arrival
rate for that day, and the normalized values for corresponding
weekdays are averaged. There are two major peaks during the day: one
at around 2~p.m. and the higher one at around
7~p.m. The higher peak
is probably due to the fact that people finish working at around
this hour and so are free to phone the call center. From 7~p.m. onward
there is a gradual decrease (except for a small increase at around
9~p.m.).
\item Fridays have a completely different pattern from the rest of the
weekdays. This can be seen in Figure \ref{fig:Scaled Intra-day
arrivals}. For each day of the week, arrivals in the 24 periods
were smoothed using the default smoothing method in the R Development Core Team (\citeyear{Rstat}). Because Friday is a half
work day for most people in Israel, it is very reasonable for its
daily pattern to differ from the rest of the weekdays.
\item
Sunday's pattern also differs from the rest of the weekdays. Figure~\ref{fig:Scaled Intra-day arrivals} exhibits how Sunday has an
earlier increase than the other weekdays (Monday through Thursday),
possibly as a result of customers who were not able to contact the
call center on the weekend (Saturday).
\end{itemize}

The above three properties lead us to consider, in Section~\ref{subsection:fixed_effects}, models with three intra-day patterns
to capture the average behavior of all six weekdays.

%

\subsection{Billing cycles}

The Private queue customers are assigned to one of four \textit{billing
cycles} when they purchase a service contract. The telecom company's
major complaint regarding their current forecasting algorithm is
that it does not incorporate billing cycles' effects. Indeed, their
own experience leads them to believe that on billing days the number
of incoming calls is higher than on nonbilling days.

Each billing cycle is defined according to two periods: the delivery
period---prior to the bank billing day, each customer receives a
letter detailing his expenses; and the billing period---the day on
which the customer's bank account will be debited. The delivery
period extends over two working days (depending solely on the
Israeli postal services). The billing period usually covers only one
day. There is usually a full week between the delivery period and
the billing period but this can vary due to weekends and holidays.
We decided therefore to describe \textit{each} cycle using two
indicators: a delivery indicator---marking the two working days of
the delivery period; and a billing indicator---marking the day of
the billing period and zero otherwise. By describing each cycle
using two indicators we actually differentiate between the influence
of the actual billing date and that of the delivery of the bill.
According to the telecom company's past experience, the different
queues are affected by different periods. For example, the Private
queue is strongly affected by the delivery periods and not so much
by the billing periods. On the other hand, the Finance queue is
strongly affected by the billing periods and less by the delivery
periods. Section~\ref{section:billingcycles} demonstrates how we
examined which indicators are significant for the Private queue's
arrival process.

\subsection{Average service times}

The various customer queues are generally served by dedicated groups
of agents. Here, we concentrate on the average service times of the
Private queue agents.

\begin{figure}[b]

\includegraphics{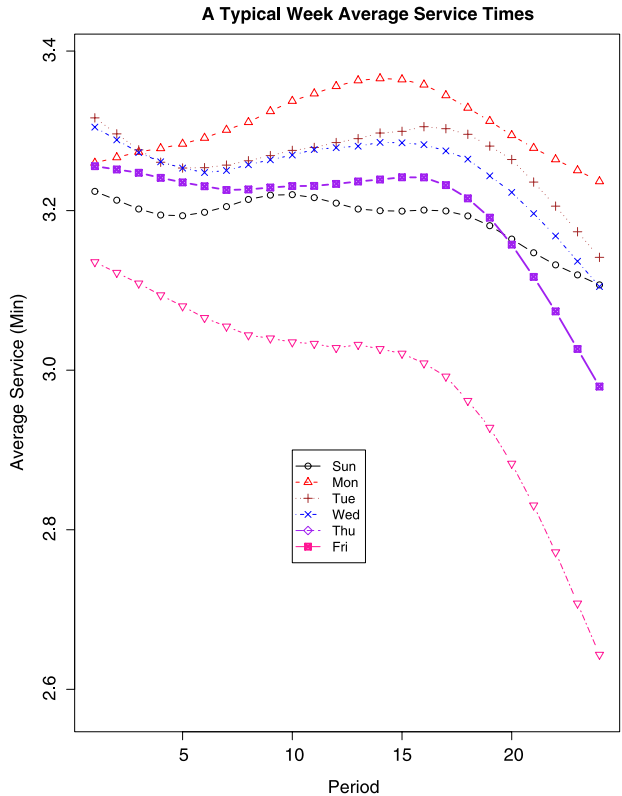}

\caption
{A typical week of average service times.}\label{fig:MeanServiceTimes}
\end{figure}

Different weekdays demonstrate different daily patterns of average
service times. Figure \ref{fig:MeanServiceTimes} demonstrates the
smoothed average service times for a typical week. Most weekdays
have a similar nonconstant pattern. Friday has a distinct
downward-sloping pattern.

%

\subsection{Staffing predictions}

The Israeli Telecom company utilizes arrivals forecast for
determining weekly staffing schedules. Each Thursday, using the past
six weeks of data as the learning data, it predicts the week
starting on Sunday ten days ahead. We will refer to this forecasting
strategy as the \textit{ten-day-ahead} weekly predictions. Accordingly,
we define three periods: the \textit{learning} period; the prediction
\textit{lead-time}, which is the duration between the last learning day
and the first predicted day; and the \textit{forecast} period.

In an attempt to mimic the challenges that the company faces, we
will predict the arrivals to the Private queue for each day between
April 11, 2004 and December~24, 2004 ($D=222$ days). For each day, we
predict the arrivals for the $K=24$ half-hour intervals between 10~a.m.
and 10~p.m., using six weeks of learning data and a lead-time period of
seven days. (In Section~\ref{lead-time} we also consider shorter
lead times and their effect on prediction accuracy.)

We have a total of $n=222\cdot24=5328$ predicted values. Excluding
the 19 irregular days (which occur during the mentioned period), we
evaluate the results using a total of 203 days or $203\cdot24=4872$
observations.

Note that although we exclude the irregular days, we preserve the
numerical distance between days in the data set by indexing them by
true dates. Later, the $\operatorname{AR}(1)$ correlation structure (in Section~\ref{random})
is fitted using the power transformation option with
true distances between days.

Our goal is to combine the arrival-process predictions with the
service-time estimates in order to predict the offered-load. We
shall then be able to estimate the staffing required in order to
achieve desired service levels.

\section{Modeling the arrival process}\label{section:arrival process}

Our approach to predicting the arrival process is to create a
parametric model for the process, then estimate its parameters based
on historical data. We discuss here the types of models considered,
how they are estimated, how they are compared and how a particular
model is ultimately chosen. Attention is also paid to justifying the
resolution (period length) used and to the dependence of the
prediction accuracy on lead-time.

\subsection{Models}
The family of models we consider allow for individual
day-of-the-week effects, period effects (possibly together with
their interactions) and exogenous effects, as well as between days
and within day (between period) dependence structures. The latter
are incorporated using random effects (or \textit{mixed} model)
formulations.

\subsubsection{Mixed model}

Let $N_{dk}$ denote the number of arrivals to the queue on day
$d=1,\ldots,D$, during the time interval $[t_{k-1},t_{k})$, where
$k=1,\ldots,K$ denotes the $k$th period of the day. Our basic
model assumption is that $N_{dk}$ follows a Poisson distribution,
with expected value $(\lambda_{dk})$. As suggested by
Brown, Zhang and Zhao (\citeyear{Rootunroot}), we take advantage of the following variance
stabilizing transformation for Poisson data: if $N_{dk}\sim
\operatorname{Poisson}(\lambda_{dk})$, then
$y_{dk}=\sqrt{N_{dk}+\frac{1}{4}}$ has approximately a mean value of
$\sqrt{\lambda_{dk}}$ and a variance of $\frac{1}{4}$. As
$\lambda_{dk} \rightarrow\infty$, $y_{dk}$ is approximately normally
distributed. In our data, $\lambda_{dk}$ has values around 500 per
half hour, hence, it is reasonable to apply this approximation.

The transformed observations $y_{dk}$ allow one to exploit the
benefits of the Gaussian linear mixed modeling approach. We model
the expected value of these observations, that is,
$\sqrt{\lambda_{dk}}$ (since their variance is known). In the mixed
model the square root of the arrival rate ($\sqrt{\lambda_{dk}}$) is
regarded as a linear function of both fixed and random effects.

Our response variable, $\mathbf{y}$, is a vector containing the
transformed observations for each period within each day, that
is, $\mathbf{y}=(y_{1,1},\ldots,y_{1,K},y_{2,1},\ldots,\break y_{D,1},\ldots,y_{D,K})^T$.
In the following paragraphs we will introduce each component of our
model, starting with the different fixed effects.

\textit{Fixed effects}: The fixed effects include the weekday
effects and the interaction between them and the period effects.
With these two effects, we capture the weekday differences in daily
levels and intra-day profiles (over the different periods).

We begin by modeling the day-level fixed effects (and later we also
discuss the intra-day level fixed effects). Formally, let ${q_d}\in
\{1,2,3,4,5,6\}$ denote the weekday which corresponds to day
$d\in\{1,\ldots,D\}$ in the data. Using this notation, introduce a
matrix $W = [w_{d,j}]$ (of dimensions $D\times6$), which is the
design matrix\footnote{ A linear model can be written as follows:
$\mathbf{Y}= \mathbf{X} \bolds{\beta}+ \bolds{\varepsilon}$. $\mathbf{X} $ is referred to as
the design matrix.} for day-of-the-week fixed-effects:
$w_{d,j}=I_{\{q_d=j\}}$, for $d\in\{1,\ldots,D\}$ and
$j\in\{1,\ldots,6\}$.

In our model we may also add exogenous variables to the above
day-level fixed effects (e.g., explanatory variables for the billing
cycles in our case study). In the next section we shall elaborate
on these additional variables but, for now, let $F_D$ denote a
$(D\times r)$ design matrix for the other explanatory day-level
variables.

Combing the two matrices, $W$ and $F_D$, produces the day-level
fixed effects design matrix, $X_D=[W,F_D]\otimes\mathbf{1}_K$, where $\mathbf{1}_K=(1,\ldots,1)^T \in\mathbb{R}^{K}$ and $\otimes$ is the
Kronecker product (often referred to as the outer product).

As mentioned before, we also consider fixed intra-day effects. Let
$X_P$ denote the corresponding design matrix. This matrix has $DK$
rows and $m$ columns. The value of $m$ can be up to $6K$ if we
assume a separate period effect for each period of each day of the
week---that is, $X_P=W \otimes I_K$, where $I_K$ is the $K \times K$
identity matrix.

\textit{Random effects}: The random effects are Gaussian
deviates with a prespecified covariance structure. One random
effect is the daily volume deviation from the fixed weekday effect.
In concert with other modeling attempts, a first-order
autoregressive covariance structure (over successive days) has been
considered for this daily deviation. It involves the estimation of
one variance parameter ($\sigma_G^2$) and one autocorrelation
parameter ($\rho_G$). Let $\gamma=(\gamma_1,\ldots,\gamma_D)^T$
denote the random day effects with covariance matrix $G$ where
$g_{i,j}= \sigma_G^2 \cdot\rho_G^{|i-j|}$. The corresponding design
matrix can be written as $Z=(I_D \otimes\mathbf{1}_K)$, where $I_D$ is
the $D$-dimensional identity matrix.

The other random effects, $\bolds{\varepsilon}$, are called the noise or
residual effects, and refer to the period-by-period random
deviations from the observed values after accounting for the fixed
weekday, period effects and the random day effects. Let $R^{*}$ be
the within-day period-by-period error covariance matrix. We assume
that this matrix is of the form $R+\sigma^2\cdot I_K$, where
$R=[r_{i,j}]$ has a first order autoregressive process covariance
structure, that is, $r_{i,j}= \sigma_R^2 \cdot\rho_R^{|i-j|}$.
Imposing this covariance structure means that if, for example, at a
certain period of the day the call center experiences a drop in the
amount of incoming calls, then we would also expect a decrease in
calls during the following periods of that day. Alternate
correlation structures for $R$ will be discussed in Section~\ref{random}.

On the basis of the square-root transformation, $\sigma^2$ should
have a value of about $0.25$.

The general formulation of our linear mixed model can be now written
as follows:
\begin{eqnarray}\label{def:mixed}
\mathbf{y} &=& X_D\bolds{\beta}_D +X_P\bolds{\beta}_P +
Z\bolds{\gamma}+\bolds{\varepsilon},
\\
\operatorname{var}(\bolds{\gamma})&=&G,\\
\operatorname{var}(\bolds{\varepsilon})&=& I_D\otimes R^{*} = I_D\otimes(R+I_k\cdot\sigma^2),\\
E(\bolds{\varepsilon})&=&\mathbf{0},\qquad  E(\bolds{\gamma})=\mathbf{0}, \qquad \bolds{\varepsilon}\bot\bolds{\gamma},
\end{eqnarray}
where:
\begin{itemize}
\item$\mathbf{y} = (y_{1,1},\ldots,y_{1,K},y_{2,1},\ldots,y_{D,1},\ldots
,y_{D,K})^T$.
\item$\beta_{D}$ is a $(6+r)$-vector of fixed effects day-level coefficients.
\item$\beta_{P}$ is a $m$-vector of fixed effects period-level coefficients.
\item$\bolds{\gamma}$ is a $(D)$-vector of random day-level
effects.
\item$\bolds{\varepsilon}$ is a $(DK)$-vector of random residuals effects.
\end{itemize}

Using the above notation, we can specify the covariance matrix of
$\mathbf{y}$ in the following manner:
\begin{equation}
V=G\otimes J_K + I_D\otimes R + I_{D\times k} \cdot\sigma^2
\qquad\mbox{where } J_K=\mathbf{1}_K \mathbf{1}^T_K.
\end{equation}

\subsubsection{The two-stage mixed model}\label{subsubsection:two stage model}

Computational problems arise when trying to implement the model in
(\ref{def:mixed}) for the data presented in Section~\ref{section:data description}.
We encountered convergence problems
when we tried to specify a parametric structure of a simple
first-order auto regression for both $R$ and $G$.

Therefore, we consider an alternative analysis based on first
modeling the daily averages in a way that is consistent with the
original model. This stage provides an estimated covariance matrix
for $G$ that later can be used when performing the second stage
analysis during which the fixed day-level and period-level effects,
as well as $R$, are estimated.

Left multiplying $\mathbf{y}$ by $M=(I_D \otimes\frac{1}{K}1^{T}_{K})$
provides a vector containing the average number of incoming calls
per period, for each day. Here is the daily-average model
formulation:
\begin{eqnarray}\label{def:daily average mixed}
M\mathbf{y} &=& [W,F_D]\bolds{\beta}_D +X^{*}_P\bolds{\beta}_P + MZ\bolds{\gamma}+M\bolds{\varepsilon},\\
\operatorname{var}(\bolds{\gamma})&=&G,\\
\operatorname{var}(\bolds{\varepsilon})&=&I_D\otimes(R+I_k \cdot\sigma^2),
\end{eqnarray}
where $X^{*}_P=M \cdot X_P$. Considering the random terms, we note
that
\begin{equation}
\operatorname{var}(M\mathbf{y}) = MVM^T = G+u\cdot I_D,
\end{equation}
where
\begin{eqnarray}
u &=& \frac{1}{K}\cdot\biggl(\frac{1}{K}\mathbf{1}_K^{T} R \mathbf{1}_K+\sigma^2 \biggr)
= \frac{1}{K} \cdot\Biggl(\frac{1}{K} \sum_{i,j}r_{ij} +\sigma^2 \Biggr).
\end{eqnarray}

Thus,\vspace*{1pt} the first stage model, as presented above in (\ref{def:daily
average mixed}), provides an estimated covariance matrix $G$, denoted
by $\hat{G}$ together with an estimate of $u$.

In the second stage we use the full model referred to in
(\ref{def:mixed}), but fix the matrix $G$ using $\hat{G}$ from the
first stage. In this stage we get estimators for all the daily fixed
effects and the periods' random effects.

The value of $\sigma^2$ should be about 0.25 if the Poisson model
holds. If the estimated value of $\sigma^2$ is close to 0.25, then we
are led to believe that the model has discovered the majority of the
predictive structure in the data, and that what is left is purely
random and unpredictable variation (at least in terms of the
underlying Poisson model). The two-stage strategy enables us to fix
$\sigma^2$ at the theoretical value of 0.2 and also enables us to
obtain estimates for this parameter. We analyze these estimated
values in Section~\ref{section:goodness}.

Another advantage of this two-stage approach is its computational
efficiency. Predicting the 203 days in our data set took
approximately 50 minutes. This makes our method valid for practical
usage since predicting a full week will take several minutes.

\subsubsection{Benchmark models}\label{subsubsection: Benchmarks}

An elementary prediction model would simply average past data for
the corresponding weekday in order to produce a forecast. This model
will be referred to as the \textit{industry model}. Denote
$W_{i,D}=\{d{}\dvtx{} d\leq i \mbox{ and } q_{D}=q_d \}$ and let $|W_{i,D}|$ denote the cardinality of $W_{i,D}$. Then the forecast
of the arrival count, $N_{D+h,k}$, utilizing the information up to
day $D$, can be expressed as
\begin{equation}\label{Equation:IndustryModel}
\hat{N}_{D+h,k}=\frac{\sum_{d \in W_{D,D+h}} N_{d,k}}{|W_{D,D+h} |}.
\end{equation}

Based on this intuitive approach, we develop two similar baseline
models. These models will serve as benchmarks for our more
complicated models.

The first basic model only considers the weekday fixed effects and
their interactions with the periods. Basically, this model states
that each day of the week has its own baseline level and its own
intra-day pattern and that consecutive days and periods are
uncorrelated [as opposed to our initial correlated mixed model
defined in (\ref{def:mixed})]. The formulation of this model can be
written as follows:
\begin{eqnarray}\label{Equation:Benchmark1}
\mathbf{y} &=& X_p \bolds{\beta}+ \bolds{\varepsilon},\\
\operatorname{var}(\bolds{\varepsilon})&=& I_{DK}\cdot(\sigma_{R}^2+\sigma^2),
\end{eqnarray}
where $\sigma^2$ has a value of about 0.25 and $X_p$ is the
intra-weekday fixed effects design matrix. This model basically
corresponds to the industry model [defined in
(\ref{Equation:IndustryModel})].

Alternatively, one can think of a different benchmark model which is
similar to the above model and also includes exogenous variables. As
a result, our second benchmark model also incorporates exogenous
variables such as the billing cycles variables. Using the above
notation, we can define the second benchmark model in the following
manner:
\begin{eqnarray}\label{Equation:Benchmark2}
\mathbf{y} &=& X_D\bolds{\beta}_D + X_P \bolds{\beta}_P + \bolds{\varepsilon},\\
\operatorname{var}(\bolds{\varepsilon})&=& I_{DK}\cdot(\sigma_{R}^2+\sigma^2),
\end{eqnarray}
where $\sigma^2$ has a value of about 0.25. The fixed effect
matrices, $X_D$ and $X_P$, have the same structures as indicated in
(\ref{def:mixed}).

This second benchmark model (similarly to the first benchmark model)
has an underlying assumption that all of its observations are
uncorrelated. Hence, it represents a baseline to our more elaborate,
correlated model.

Both of the benchmark models are, in fact, linear regression
models and are quite fast and efficient in providing the necessary
predictions using standard programs such as R [R Development Core Team (\citeyear{Rstat})] and
SAS\tsup{\textregistered}/STAT [\cite{SAS}].

\subsection{Model evaluation criteria}\label{Evaluation}
Suppose now that we aim to predict arrivals over $D^P$ days, for
each of the $K$ periods in those days; altogether $n=D^PK$ predicted
values. Let $\hat{N}_{dk}$ denote the predicted value of ${N}_{dk}$,
which is the number of arrivals in the $k$th period for
day $d$. We define two measures to compare between the observed and
the predicted values:
\begin{itemize}
\item Squared Error: $\mathrm{SE}_{dk}=(\hat{N}_{dk}-{N}_{dk})^2$,
\item Relative Error: $\mathrm{RE}_{dk}=100\cdot\frac{|\hat{N}_{dk}-{N}_{dk}|}{{N}_{dk}}$.
\end{itemize}

The following two measures are used to evaluate confidence
statements concerning $N_{dk}$:
\begin{itemize}
\item $\operatorname{Cover}_{dk} =\mathbf{I}(N_{dk}\in(\operatorname{Lower}_{dk},\operatorname{Upper}_{dk}))$,
\item $\operatorname{Width}_{dk} = \operatorname{Upper}_{dk} - \operatorname{Lower}_{dk}$.
\end{itemize}

In the above, Lower$_{dk}$ and Upper$_{dk}$ denote the lower and
upper (nominally 95\%) confidence limits. We obtain these limits
using conditional multivariate Gaussian theory. We can use this
theory since we assume that the observations $y_{dk}$ follow the
Gaussian distribution. For further details on how these limits are
computed, the reader is referred to Henderson (\citeyear{BLUP}).

The comparison between different forecasting models is performed
over the entire set of $n$ predicted values or \textit{out-of-sample}
predictions. For each day we predict the arrival counts for each of
the twenty-four periods based on six weeks of past data and a
lead-time of one week. This procedure is carried out 203 times since
there are 203 regular weekdays between April 11, 2004 and December
25, 2004. Then we average the following measures for each day over
the $K$ periods:
\begin{itemize}
\item RMSE$_{d}=\sqrt{\frac{\sum_{k=1}^{K}\mathrm{SE}_{dk}}{K}}$,
\item APE$_{d}=\frac{\sum_{k=1}^{K}\mathrm{RE}_{dk}}{K}$,
\item Cover$_{d}=\frac{\sum_{k=1}^{K}\operatorname{Cover}_{dk}}{K}$,
\item Width$_{d}=\frac{\sum_{k=1}^{K}\operatorname{Width}_{dk}}{K}$.
\end{itemize}

This procedure results in 203 daily values for each of these
measures. We then report for each measure the lower quartile, the
median, the mean and the upper quartile values of these daily
summary statistics.

These four measures reflect different aspects of prediction accuracy.
The RMSE represents an average prediction
error, while the APE is the average percentage error. They both give
a sense of how well our point estimates are performing. The coverage
probability and the prediction interval width relate to prediction
intervals. The width of the prediction intervals are constructed
using a nominal confidence level of $95\%$. Hence, we expect the
mean coverage probability to be close to $95\%$. A model which
performs well will have a narrow prediction width with coverage probabilities
close to $95\%$.

\subsection{Choosing the model}
In practice, there may be several proposed exogenous variables that
may have an effect on the call arrival process. Our strategy for
screening those variables is to first work with models at the daily
level, that is, to find those exogenous variables that have an impact
on the daily counts. It is typically too burdensome to use a full
mixed model analysis in order to screen for exogenous day-level
variables.

After having chosen the day-level exogenous variables, the next step
is to fit full mixed models and to evaluate different covariance
structures at both the between-day and the within-day
(between-period) levels. For further discussion on the theory of
mixed models, the reader is referred to Demidenko (\citeyear{Mixed}).

We illustrate these steps for our case study data.

\subsubsection{Exogenous variables}\label{section:billingcycles}

As mentioned in the data description, in our case study there are
eight indicators which represent the four major billing cycles
(i.e., four delivery period indicators and four billing period
indicators). Based on the company's information, we were made aware
that some of these indicators might not have a significant influence
on the Private queue's arrival process. The purpose of this
preliminary examination is to highlight those indicators which are
significant so they may later be incorporated in the final
forecasting model. For this coarse investigation, we use the
aggregated \textit{daily} arrivals between February
$14$th, 2004 and December 31st, 2004
(excluding all twenty-two outliers).

Let $N_{d}$ denote the number of arrivals to the Private queue on
day $d=1,\ldots,D$. As before, let ${q}_{d}$ denote the weekday
corresponding to day $d$. The daily arrivals are modeled using a
Poisson log-linear model [for further details on this approach, the
reader is referred to McCullagh and Nedler (\citeyear{GLM})]. Our initial model is of the
following form:
\begin{eqnarray}\label{model:genmod}
N_{d} &\sim& \operatorname{Poisson}(\lambda_{d}), \\
\nonumber\log(\lambda_{d}) &=&
\sum_{l=1}^{6}\mathrm{W}_{l} \cdot X^l_{q_{d}}
+\sum_{j\in\mathrm{cycles}}\mathrm{B}_{j} \cdot
\mathrm{P}^{j}_{d}+\sum_{j\in\mathrm{cycles}}\mathrm{D}_{j}\cdot
\mathrm{U}^{j}_{d},
\end{eqnarray}
where
\begin{eqnarray*}\label{def:genmod}
&&\hspace*{-27pt}\mathrm{W}_{l} \mbox{ is the $l$th weekday coefficient,}\\
{X_{q_d}^l} &=&
\cases{
1, &\quad if $l=q_{d}$,\cr
0, &\quad otherwise,
}
\\
&&\hspace*{-25pt}\mathrm{B}_{j} \mbox{ is the $j$th billing period coefficient,} \\
\mathrm{P}^{j}_{d} &=&
\cases{1, &\quad if cycle $j$'s billing period falls on the $d$th day,\cr
0, &\quad otherwise,}
\\
&&\hspace*{-26pt}\mathrm{D}_{j} \mbox{ is the $j$th delivery period coefficient,} \\
\mathrm{U}^{j}_{d} &=&
\cases{1, &\quad if cycle $j$'s delivery period falls on the $d$th day,\cr
0, &\quad otherwise.}
\end{eqnarray*}

The model may be implemented using the \textit{GENMOD} procedure
in\break
SAS\tsup{\textregistered}/STAT [Aldor-Noiman, Feigin and Mandelbaum (\citeyear{supp})] based on the 254 observations in
the current learning set. Some of the results are summarized in
Table~\ref{table:genmod parameters}.

The results indicate the following: the six weekdays have
significantly different effects, each having a different baseline
mean; the \textit{delivery} period indicators are significant and
have a positive effect on the mean value of the number of incoming
calls; on the other hand, most of the \textit{billing} period
indicators seem less significant, which confirms the telecom
company's beliefs. The Cycle 14 billing period seems to have an
exceptional effect. First, it is statistically significant as
opposed to the billing indicators of the rest of the cycles.
Furthermore, its estimator is the only negative value among those
of all of the effects. This noticeable result would suggest that
the number of incoming calls is reduced during the Cycle 14
billing period. We could not attribute this phenomenon to any
outlying data problems. One possibility is that this negative
value can be compensating for other oversized billing cycle
effects, which could arise due to the overlap of delivery and
billing days among the 4 cycles.

%
\begin{table}
\tabcolsep=6pt
\caption
{Analysis of parameter estimates for the Poisson log-linear model}\label{table:genmod parameters}
\begin{tabular*}{\textwidth}{@{\extracolsep{\fill}}lcd{2.4}d{2.2}d{2.4}@{}}
\hline
\textbf{Parameter} & \textbf{Category} & \multicolumn{1}{c}{\textbf{Estimate}}
& \multicolumn{1}{c}{\textbf{Chi-square}} & \multicolumn{1}{c@{}}{$\mathbf{Pr>ChiSq}$}
\\
\hline
$W_{q}$ & Sunday & 9.5358 & \multicolumn{1}{c}{718{,}177} & < 0.0001
\\
$W_{q}$ & Monday & 9.4977 & \multicolumn{1}{c}{541{,}484} & < 0.0001
\\
$W_{q}$ & Tuesday & 9.4880 & \multicolumn{1}{c}{572{,}102} & < 0.0001
\\
$W_{q}$ & Wednesday & 9.4719 & \multicolumn{1}{c}{649{,}509} & < 0.0001
\\
$W_{q}$ & Thursday & 9.4326 & \multicolumn{1}{c}{677{,}894} & < 0.0001
\\
$W_{q}$ & Friday & 9.0385 & \multicolumn{1}{c}{453{,}474} & < 0.0001
\\[1pt]
D$^{1}$ & $\cdot$ & 0.0586 & 11.59 & 0.0007
\\
D$^{7}$ & $\cdot$ & 0.0327 & 3.71 & 0.0540
\\
D$^{14}$ & $\cdot$ & 0.0449 & 5.39 & 0.0202
\\
D$^{21}$ & $\cdot$ & 0.0935 & 17.98 & < 0.0001
\\
B$^{1}$ & $\cdot$ & 0.0242 & 2.10 & 0.1473
\\
B$^{7}$ & $\cdot$ & 0.0279 & 2.17 & 0.1403
\\
B$^{14}$ & $\cdot$ & -0.0592 & 6.83 & 0.0090
\\
B$^{21}$ & $\cdot$ & 0.0276 & 2.54 & 0.1109
\\\hline
\end{tabular*}
\end{table}

These results led us to believe that some of the explanatory
variables are statistically redundant. We proceeded by comparing
different models with this initial model [defined in
(\ref{model:genmod})] using the ``contrast'' statement in the
\textit{GENMOD} procedure (which computes likelihood-ratio
statistics). The different models were variations of the initial
model. They excluded different covariates in order to establish the
significance of the omitted variables.

The results indicated that billing periods 1, 7 and 21 are
redundant. It is therefore clear that there are three main factors
contributing to the daily volumes: the weekday, the delivery periods
and billing period of Cycle 14. This analysis led to two possible
modeling alternatives: (i) The first model includes the weekday
effect, Cycle 14 billing period indicator and the four delivery
period indicators; (ii)~The second model includes the weekday
effect, Cycle 14 billing period indicator and the one global
delivery period indicator (which takes the value one when at least
one of the cycles is during its delivery period). The results of the
two models are shown in Table \ref{table:genmod contrast}. In the
next section we will explore both of these setups and choose the one
which yields better out-of-sample results.

%
\begin{table}
\caption
{Log-linear models contrasts analyses. Each row depicts a different model which is
compared to the initial model. Comparing models \textup{1} and \textup{2} to the
initial model shows that the extra variables are not significant at
a significance level of 5$\%$}\label{table:genmod contrast}
\begin{tabular*}{\textwidth}{@{\extracolsep{\fill}}llcc@{}cc@{}}
\hline
&& \multirow{3}{43pt}{\centering\textbf{Numerator degrees of freedom}}
& \multirow{3}{54pt}{\centering\textbf{Denominator degrees of freedom}}
\\
&\multirow{2}{114pt}{\centering\textbf{The alternative model explanatory variables}}
\\
\textbf{Model no.}& & & & $\bolds{F}$ & $\bolds{\mathrm{Pr}>F}$ \\
\hline
1 & Weekday, &6 & 240 & 1.89 & 0.0834 \\
& billing 14 period indicator, & & & & \\
& global delivery period indicator & & & &\\[3pt]
2 & Weekday, & 3 & 240 & 2\phantom{.89} &0.1152 \\
& billing 14 period indicator, & & & &\\
& four delivery period indicators & & & &\\
\hline
\end{tabular*}
\end{table}

\subsubsection{Adjusting the fixed
effects}\label{subsection:fixed_effects}

Our process of model selection does not rely on classical inference
methods or measures, such as $F$-test or Akaike's Information Criteria
[Sakamoto, Ishiguro and Kitagawa (\citeyear{AIC})]. These methods rely on the learning-set and its
dimension but do not consider the models forecasting qualities.

Therefore, we decide to explore the influence of the models'
elements on the prediction performance based on the 2004
validation-set. The evaluation criteria were detailed in Section~\ref{Evaluation}.

Keeping the parsimony\footnote{Parsimony refers to the
philosophic rule where the simplest of competing theories/models
be preferred to the more complex.} concept in mind, Figure~\ref{fig:Scaled Intra-day arrivals} suggests that some weekday
patterns resemble others. Mainly, Monday through Thursday have
similar intra-day period profiles. We shall explore two
alternatives for the weekday pattern setting: (i) a model which
includes a different period profile for each weekday---we refer
to this setting as the \textit{Multi-Pattern} setting; (ii) a
model which includes three intra-day patterns (i.e., one for
Sunday, one for Friday and one for the rest of the weekdays)---we refer to this setting as the \textit{Three-Pattern} setting.

Combining the two different settings for the intra-day effects with
the two settings of the billing cycles effects (detailed in Section~\ref{section:billingcycles}) results in four different models: (i)
Model 1---is a Three-Pattern model which also includes Cycle 14
billing period and one global delivery period indicator; (ii) Model~2---is a Three-Pattern model which also includes Cycle 14 billing
period and four delivery period indicators; (iii) Model 3---is a
Multi-Pattern model which also includes Cycle 14 billing period and
one global delivery period indicator; (iv) Model 4---is a
Multi-Pattern model which also includes Cycle 14 billing period and
four delivery period indicators.

For now, we shall set the between-periods (within-day) correlation
structure according to a first-order autoregressive structure. We
choose this specific structure for its simplicity. In the next
section we will also consider other correlation structures. We also
fix the value of $\sigma^{2}$ to its 0.25 theoretical value.

In addition, we compare the four models with the performance of the
two benchmark models, mentioned in Section~\ref{subsubsection:
Benchmarks}. The first benchmark model only includes the weekday
effect and the weekday period profiles. The alternative benchmark
model has the Multi-Pattern fixed effects and two additional billing
cycles' indicators: one global delivery indicator and the billing
period indicator associated with Cycle 14. The reason why these
specific fixed effects settings were chosen will be explained later
in this section.

These six models are evaluated using the same out-of-sample
prediction procedure previously discussed.

%
\begin{table}
\caption
{RMSE results for the three fixed effects models and two benchmarks models}\label{Table:4 Fixed models RMSE Results}
\begin{tabular*}{\textwidth}{@{\extracolsep{\fill}}lccccc@{}}
\hline
& \multicolumn{5}{c@{}}{\textbf{RMSE}}
\\[-6pt]
& \multicolumn{5}{c@{}}{\hrulefill}\\
$\bolds{N=203}$ & \textbf{Model 1} & \textbf{Model 2} & \textbf{Model 3} & \textbf{Benchmark 1} & \textbf{Benchmark 2}\\
\hline
1st quartile & 31.80 & 32.28 & 32.55 & 33.24 & 32.39\\
Median & 39.30 & 40.16 & 40.19 & 40.54 & 39.98\\
Mean & 44.78 & 47.19 & 45.49 & 46.51 & 45.45\\
3rd quartile & 51.63 & 58.29 & 52.38 & 54.88 & 53.16 \\
\hline
\end{tabular*}
\end{table}

%
\begin{table}[b]
\caption
{APE results for the four fixed
effects models and two benchmarks models}\label{Table:4 Fixed
models APE Results}
\begin{tabular*}{\textwidth}{@{\extracolsep{\fill}}ld{2.2}d{2.2}d{2.2}d{2.2}d{2.2}@{}}
\hline
& \multicolumn{5}{c@{}}{\textbf{APE}} \\[-6pt]
& \multicolumn{5}{c@{}}{\hrulefill}\\
\multicolumn{1}{@{}l}{$\bolds{N=203}$} & \multicolumn{1}{c}{\textbf{Model 1}}
& \multicolumn{1}{c}{\textbf{Model 2}} & \multicolumn{1}{c}{\textbf{Model 3}}
& \multicolumn{1}{c}{\textbf{Benchmark 1}}
& \multicolumn{1}{c@{}}{\textbf{Benchmark 2}}\\
\hline
1st quartile & 6.51 & 6.57 & 6.74 & 6.81 & 6.77\\
Median & 8.13 & 8.45 & 8.12 & 8.24 & 8.09 \\
Mean & 9.27 & 9.75 & 9.41 & 9.58 & 9.41\\
3rd quartile & 11.27 & 11.75 & 11.34 & 11.29 & 11.31
\\ \hline
\end{tabular*}
\end{table}

We use the SAS\tsup{\textregistered}/STAT \textit{Mixed} procedure in order to
implement and evaluate the candidate models. Convergence problems
occurred when we tried to implement Model 4 with the Multi-Pattern
and the billing cycles effects. Therefore, this model was
subsequently excluded from the analysis.

Tables~\ref{Table:4 Fixed models RMSE Results}, \ref{Table:4 Fixed
models APE Results}, \ref{Table:4 Fixed models Cover Results} and
\ref{Table:4 Fixed models Width Results} present the results of the
three different fixed model setups. Out of the three different
models, the 1st alternative seems to exhibit the best
results. Its results are generally better than the benchmark models.
One might argue that the shorter confidence intervals imply that the
benchmark models outperform the first model. However, since their
coverage probabilities are very far from the nominal $95\%$, we
conclude that these narrow intervals are unreliable and probably
result from an underestimated error variance, that is, $\sigma_R^2$.
This phenomenon can occur when not all of the sources of variability
are accounted for in the model and, in particular, when correlation
structure in the data is ignored.

%
\begin{table}
\caption
{Coverage probabilities for the four fixed effects models and two benchmarks models}\label{Table:4 Fixed models Cover Results}
\begin{tabular*}{\textwidth}{@{\extracolsep{\fill}}lccccc@{}}
\hline
& \multicolumn{5}{c@{}}{\textbf{Coverage probability}} \\[-6pt]
& \multicolumn{5}{c@{}}{\hrulefill}\\
$\bolds{N=203}$ & \textbf{Model 1} & \textbf{Model 2} & \textbf{Model 3} & \textbf{Benchmark 1} & \textbf{Benchmark 2}\\
\hline
1st quartile & 0.92 & 0.88 & 0.92 & 0.33& 0.33\\
Median & 0.96 & 0.96 & 0.96 & 0.50& 0.50 \\
Mean & 0.93 & 0.92 & 0.93 & 0.49& 0.49\\
3rd quartile & 1\phantom{.92} & 1\phantom{.92} & 1\phantom{.92} & 0.63& 0.63\\ \hline
\end{tabular*}
\end{table}

%
\begin{table}[b]
\caption
{Confidence interval widths for the three fixed effects models and two
benchmarks models}\label{Table:4 Fixed models Width Results}
\begin{tabular*}{\textwidth}{@{\extracolsep{\fill}}lccccc@{}}
\hline
& \multicolumn{5}{c@{}}{\textbf{Width}} \\[-6pt]
& \multicolumn{5}{c@{}}{\hrulefill}\\
$\bolds{N=203}$ & \textbf{Model 1} & \textbf{Model 2} & \textbf{Model 3} & \textbf{Benchmark 1} & \textbf{Benchmark 2}\\
\hline
1st quartile & 150.03 & 149.14 & 149.13 & 52.62 & 51.44\\
Median & 168.99 & 169.22 & 168.81 & 59.67 & 58.80\\
Mean & 175.47 & 178.24 & 174.31 & 62.92 & 61.44\\
3rd quartile & 195.27 & 194.61 & 197.81 & 68.95 & 67.86
\\ \hline
\end{tabular*}
\end{table}

The second benchmark model includes the same billing cycles' settings
as the first model. It can be thought of as a ``fixed'' version of the
first model, that is, without the random effects. It is interesting
to see that introducing the intra- and inter-day correlations
generally improves the forecasting results. Intuitively speaking,
the additional correlation parameters result in wider confidence
intervals to compensate for the extra uncertainty.

Following these analyses, the models considered in the next
subsection will all have the fixed effects of the 1st
model. These fixed effects include the Three-Pattern intra-day
effect and the two indicators of the billing cycles.

\subsubsection{Dependence structures}\label{random}

Having chosen the fixed effects that will be incorporated in our
model, we now discuss the modeling of the random effects. There are
two sources of variation in our model: one is from the daily volume
effect $\gamma$ and the other is the within-day error vector
$\varepsilon$.

We begin by examining different structures for the matrix $R$ which
is the within-day covariance matrix where the residuals' variance
(i.e., $\sigma^{2}$) is confined to its theoretical value of 0.25.
Based on previous research [e.g., Avramidis, Deslauriers and L'Ecuyer (\citeyear{avramidis-et-al})],
we consider two simple time series structures for $R$ which both can
account for the strong intra-day correlations. The first is an $\operatorname{AR}(1)$
(first order auto-regressive) implying that $r_{i,j}=\sigma_R^2
\cdot\rho^{|i-j|}$. The second is an $\operatorname{ARMA}(1,1)$ (auto-regressive
moving averages) which means that
$r_{i,j}= \sigma_R^2 \cdot\delta \cdot\rho^{|i-j|-1}$ (the covariance between periods $i$ and $j$ where
$i \neq j$) and $r_{i,i}=\sigma_R^2$ (which is the variance of
period $i$). Other covariance structures theoretically may also be
incorporated here, but since most of them are more complex (such as
the Toeplitz form, which includes more parameters), we did not
consider them because of computational limitations. Note also that
most other forms of covariance matrices in SAS\tsup{\textregistered}/STAT are
not directly related to time series structures.

We evaluate and compare the models using the same technique we
developed for selecting the fixed effects. We use the same
learning data as before.

The results are shown in Tables \ref{Table: within-day error
covariance structure RMSE}, \ref{Table: within-day error covariance
structure APE}, \ref{Table: within-day error covariance structure
cover} and \ref{Table: within-day error covariance structure width}.
The results for the $\operatorname{ARMA}(1,1)$ show only slight improvements in the
coverage probability compared to the $\operatorname{AR}(1)$ structure. However, it
seems that the point predictions are slightly better using the $\operatorname{AR}(1)$
structure. Another factor that should be taken into consideration is
that the CPU time was higher for the $\operatorname{ARMA}(1,1)$ model. In conclusion,
the model chosen for $R$ in this approach is the $\operatorname{AR}(1)$ model for the
residual error vector.

%
\begin{table}
\tablewidth=7cm
\caption
{Different within-day errors covariance structure. RMSE results}\label{Table: within-day error covariance
structure RMSE}
\begin{tabular*}{\tablewidth}{@{\extracolsep{\fill}}lcc@{}}
\hline
$\bolds{N=203}$ & \multicolumn{2}{c@{}}{\textbf{RMSE}} \\[-6pt]
& \multicolumn{2}{c@{}}{\hrulefill} \\
\textbf{\textit{R} covariance structure} & $\bolds{\operatorname{AR}(1)}$ & $\bolds{\operatorname{ARMA}(1,1)}$ \\
\hline
1st quartile & 31.80 & 31.85 \\
Median & 39.30 & 39.36 \\
Mean & 44.78 & 44.81 \\
3rd quartile & 51.63 & 51.26 \\ \hline
\end{tabular*}
\end{table}

%
\begin{table}[b]
\tablewidth=7cm
\caption
{Different within-day errors covariance
structure.\break APE results}\label{Table: within-day error covariance
structure APE}
\begin{tabular*}{\tablewidth}{@{\extracolsep{\fill}}lcc@{}}
\hline
$\bolds{N=203}$ & \multicolumn{2}{c@{}}{\textbf{APE}} \\[-6pt]
& \multicolumn{2}{c@{}}{\hrulefill} \\
\textbf{\textit{R} covariance structure} & $\bolds{\operatorname{AR}(1)}$ & $\bolds{\operatorname{ARMA}(1,1)}$ \\
\hline
1st quartile \phantom{0}& 6.51 & \phantom{0}6.52 \\
Median & \phantom{0}8.13 & \phantom{0}8.12 \\
Mean & \phantom{0}9.27 & \phantom{0}9.28 \\
3rd quartile & 11.27 & 11.20 \\ \hline
\end{tabular*}
\end{table}

%
\begin{table}
\tablewidth=7cm
\caption
{Different within-day errors covariance structure comparison. Coverage
results}\label{Table: within-day error covariance structure cover}
\begin{tabular*}{\tablewidth}{@{\extracolsep{\fill}}lcc@{}}
\hline
$\bolds{N=203}$ & \multicolumn{2}{c@{}}{\textbf{Coverage probability}} \\[-6pt]
& \multicolumn{2}{c@{}}{\hrulefill} \\
\textbf{\textit{R} covariance structure} & $\bolds{\operatorname{AR}(1)}$ & $\bolds{\operatorname{ARMA}(1,1)}$ \\
\hline
1st quartile & 0.92 & 0.92 \\
Median & 0.96 & 1\phantom{.92} \\
Mean & 0.93 & 0.94 \\
3rd quartile & 1\phantom{.92} & 1\phantom{.92} \\ \hline
\end{tabular*}
\end{table}

%
\begin{table}[b]
\tablewidth=7cm
\caption
{Different
within-day errors covariance structure.\break Width results}\label{Table: within-day error covariance structure width}
\begin{tabular*}{\tablewidth}{@{\extracolsep{\fill}}lcc@{}}
\hline
$\bolds{N=203}$ & \multicolumn{2}{c@{}}{\textbf{Width}} \\[-6pt]
& \multicolumn{2}{c@{}}{\hrulefill} \\
\textbf{\textit{R} covariance structure} & $\bolds{\operatorname{AR}(1)}$ & $\bolds{\operatorname{ARMA}(1,1)}$ \\
\hline
1st quartile & 150.03 & 154.93 \\
Median & 168.99 & 174.01 \\
Mean & 175.47 & 179.96 \\
3rd quartile & 195.27 & 204.09 \\ \hline
\end{tabular*}
\end{table}

The last source of variability is the daily volume effect, $G$. We
assume that its covariance structure also has a first-order
autoregressive form. This basic assumption means that if on a
certain day the call center experienced a rise in the amount of
incoming calls (compared to the fixed effects prediction), then we
would also expect to see a similar increase during the following
days. As the days become farther apart from that day, we expect its
influence to decline.

We investigated the influence of the $\gamma$ correlations by
comparing our mixed model with an alternative model which does not
include this random effect. Using the company's current strategy for
prediction which includes a relatively long lead-time of one week,
one would hardly expect to see any difference between the two
models. The results are summarized in Tables~\ref{Table: G
covariance structure RMSE}, \ref{Table: G covariance structure APE},
\ref{Table: G covariance structure cover} and \ref{Table: G
covariance structure width}. The results show a slight improvement
after incorporating an $\operatorname{AR}(1)$ structure.

To examine the influence of the between-day random effect, we
further explore the results of the above two models by changing the
lead-time period. This analysis is presented in Section~\ref{lead-time}. The analysis reveals that the model incorporating
the $\operatorname{AR}(1)$ inter-day covariance structure produces better results.

%
\begin{table}
\caption
{Testing the influence of the daily random effect. RMSE results}\label{Table: G covariance structure RMSE}
\begin{tabular*}{10cm}{@{\extracolsep{\fill}}lcc@{}}
\hline
& \multicolumn{2}{c@{}}{\textbf{RMSE}} \\[-6pt]
& \multicolumn{2}{c@{}}{\hrulefill} \\
& \multirow{2}{80pt}{\centering\textbf{Model with} $\bolds{\operatorname{AR}(1)}$ \textbf{daily random effect}}
& \multirow{2}{80pt}{\centering\textbf{Model without a daily random effect}}\\
$\bolds{N=203}$ & \\
\hline
1st quartile & 31.80 & 31.92 \\
Median & 39.30 & 39.67 \\
Mean & 44.78 & 44.83 \\
3rd quartile & 51.63 & 50.40 \\ \hline
\end{tabular*}
\end{table}

%
\begin{table}[b]
\caption
{Testing the influence of the daily random effect. APE results}\label{Table: G covariance structure APE}
\begin{tabular*}{10cm}{@{\extracolsep{\fill}}lcc@{}}
\hline
& \multicolumn{2}{c@{}}{\textbf{APE}} \\[-6pt]
& \multicolumn{2}{c@{}}{\hrulefill} \\
& \multirow{2}{80pt}{\centering\textbf{Model with} $\bolds{\operatorname{AR}(1)}$ \textbf{daily random effect}}
& \multirow{2}{80pt}{\centering\textbf{Model without a daily random effect}}\\
$\bolds{N=203}$ & \\
\hline
1st quartile & \phantom{0}6.51 & \phantom{0}6.46 \\
Median & \phantom{0}8.13 & \phantom{0}8.15 \\
Mean & \phantom{0}9.27 & \phantom{0}9.28 \\
3rd quartile & 11.27 & 11.09 \\ \hline
\end{tabular*}
\end{table}

\subsubsection{Final model choice---goodness of fit}\label{section:goodness}

\begin{figure}

\includegraphics{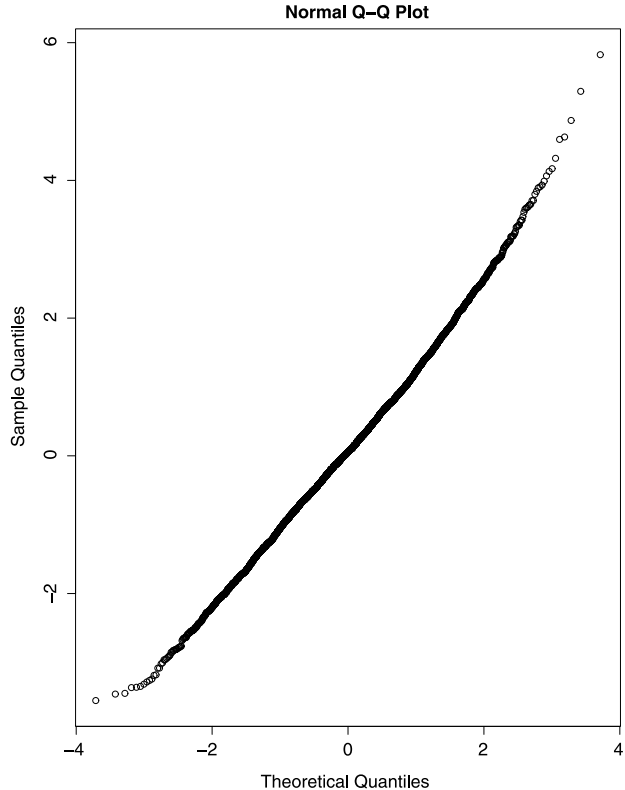}

\caption
{The model residuals QQ plot.}\label{fig:Mixed qqplot}
\end{figure}

Our final model includes an intra-day pattern for each weekday, two
billing cycles indicators and a first order auto-regressive
structure
for both the intra-day and the inter-day correlations. Figure~\ref{fig:Mixed qqplot} presents the QQ-plot for the residuals of our
final model. It appears that the assumption of normally distributed
residuals is reasonable.

As mentioned in previous sections, the theoretical value of
$\sigma^2$, described in (\ref{def:mixed}), is 0.25. We reran our
chosen model, but this time we did not specify the value of
$\sigma^2$ in advance. Instead we computed the
estimated value of
this parameter. Since we predict 203 days in our database (each time
based on 6 weeks data with a lead-time of 7 days), we can obtain 203
values of this parameter. Based on these values, the mean estimated
value was 0.29. Since the estimated value of $\sigma^2$ is close to
0.25, we are inclined to believe that the model has discovered the
majority of the predictive structure in the data, and that what is
left is purely random and unpredictable variation.

\subsection{Justifying the half-hour periods}\label{section:halfhour}

An interesting debate might be held between practitioners and
theoreticians as to what is the appropriate interval resolution to
analyze. Theoreticians might say that in order to fully maintain the
homogeneity assumption the intervals should be as small as possible.
Alternatively, from a practitioner's point of view, the resolution
should be determined as a function of managerial flexibility: for
example, if it is feasible to change the number of available agents
every 5 minutes, then this should be the appropriate interval
resolution. (Such high flexibility can occur, e.g., in large
call centers where there are typically agents who are occupied with
various off-line tasks and who can be made immediately available for
service.) However, our experience suggests that call centers
commonly plan their daily schedule according to either half-hour or
15-minute resolutions.

%
\begin{table}
\caption
{Testing the influence of the daily random effect.
Coverage results}\label{Table: G covariance structure cover}
\begin{tabular*}{10cm}{@{\extracolsep{\fill}}lcc@{}}
\hline
& \multicolumn{2}{c@{}}{\textbf{Coverage probability}} \\[-6pt]
& \multicolumn{2}{c@{}}{\hrulefill} \\
& \multirow{2}{80pt}{\centering\textbf{Model with} $\bolds{\operatorname{AR}(1)}$ \textbf{daily random effect}}
& \multirow{2}{80pt}{\centering\textbf{Model without a daily random effect}}\\
$\bolds{N=203}$ & \\
\hline
1st quartile & 0.92 & 0.88 \\
Median & 0.96 & 0.96 \\
Mean & 0.93 & 0.92 \\
3rd quartile & 1.00 & 1.00 \\ \hline
\end{tabular*}
\end{table}

%
\begin{table}[b]
\caption
{Testing the influence of the daily random effect. Width results}\label{Table: G covariance structure width}
\begin{tabular*}{10cm}{@{\extracolsep{\fill}}lcc@{}}
\hline
& \multicolumn{2}{c@{}}{\textbf{Width}} \\[-6pt]
& \multicolumn{2}{c@{}}{\hrulefill} \\
& \multirow{2}{80pt}{\centering\textbf{Model with} $\bolds{\operatorname{AR}(1)}$ \textbf{daily random effect}}
& \multirow{2}{80pt}{\centering\textbf{Model without a daily random effect}}\\
$\bolds{N=203}$ & \\
\hline
1st quartile & 150.03 & 144.20 \\
Median & 168.99 & 163.49 \\
Mean & 175.47 & 165.95 \\
3rd quartile & 195.27 & 186.43 \\ \hline
\end{tabular*}
\end{table}

Our Gaussian mixed model can be easily modified to deal with
different interval resolutions. An interesting question is how much
worse are predictions based on lower-level resolutions than 15-minute
predictions, when evaluated at the 15-minute period level.
For example, if one predicted accurately the arrival count over a
half-hour period, but in that period the first 15 minutes had 0.5
times the average arrival rate, and the second 15 minutes had 1.5
times the average arrival rate, then using the half-hour prediction
(by dividing it equally over each 15-minute interval) would lead one
to seriously overstaff during the first 15 minutes and understaff
during the second 15 minutes. This problem would not have happened
if one had good predictions at the 15-minute resolution.

Hence, we are interested in analyzing the effect of the interval
resolution on the forecast accuracy at the finest practical
resolution.

To this end, we use our Israeli Telecom data and predict the
arrival counts between 10~a.m. and 10~p.m. during the 203 regular
weekdays between April 11 and December 24, 2004. The forecast
procedure is the same as before, meaning that we used 6 weeks of
past data as learning data to predict the day which begins seven
days ahead.

Our baseline data resolution is 15-minute intervals. We compared
15-minute intervals with three additional interval resolutions: \textit{half-hour}, \textit{one-hour} and \textit{four-hours}. In order to fairly assess the
behavior of the different interval widths, we scaled the predictions
into 15-minute blocks: the lower-resolution forecasts were simply
\textit{equally} distributed among the 15-minutes intervals. For
example, we took the predicted arrival count for a specific hour (on
a certain day) and equally divided it over its four quarter hours.

%

Tables \ref{Table:Resolution analysis RMSE Results} and
\ref{Table:Resolution analysis APE Results} describe the results for
both the RMSE and APE, respectively. Using the RMSE measure, we see
that the half-hour resolution predictions are virtually as precise
as the 15-minute ones, and somewhat better than the one-hour
resolution predictions. However, using the RMSE measure puts the
forecasts based on wider intervals at a disadvantage and, thus, we
also take a look at the APE results.

%
\begin{table}
\caption
{Prediction accuracy as a function of interval resolution. We
compare the RMSE result of the mixed model for four different
resolutions: 15-minute, half-hour, one-hour and four-hour}\label{Table:Resolution analysis RMSE Results}
\begin{tabular*}{\textwidth}{@{\extracolsep{\fill}}lcccc@{}}
\hline
& \multicolumn{4}{c@{}}{\textbf{RMSE}} \\[-6pt]
&\multicolumn{4}{c@{}}{\hrulefill} \\
& \textbf{15-minutes} & \textbf{Half-hour} & \textbf{One-hour }& \textbf{Four-hour} \\ \hline
Min & 12.28& 12.07 & 14.13 & 32.18 \\
1st quartile & 19.67 & 19.81 & 20.86 & 38.18\\
Median & 22.48 & 22.59 & 23.56 & 41.37\\
Mean & 24.97 & 25.06 & 25.87 & 42.80 \\
3rd quartile & 28.45 & 28.29 & 29.18 & 46.14 \\
Max & 60.00 & 60.01 & 60.21 & 73.12
\\ \hline
\end{tabular*}
\end{table}

%
\begin{table}[b]
\caption
{Prediction accuracy comparison as a
function of interval resolution. We compare the APE result of the
mixed model with four different resolutions: 15-minute, half-hour,
one-hour and four-hour}\label{Table:Resolution analysis APE
Results}
\begin{tabular*}{\textwidth}{@{\extracolsep{\fill}}lcccc@{}}
\hline
& \multicolumn{4}{c@{}}{\textbf{APE}} \\[-6pt]
&\multicolumn{4}{c@{}}{\hrulefill} \\
& \textbf{15-minutes} & \textbf{Half-hour} & \textbf{One-hour} & \textbf{Four-hour} \\ \hline
Min & \phantom{0}5.68 & \phantom{0}5.57 & \phantom{0}5.36 & \phantom{0}7.17 \\
1st quartile & \phantom{0}8.01 & \phantom{0}8.03 & \phantom{0}8.14 & \phantom{0}9.74 \\
Median & \phantom{0}9.30 & \phantom{0}9.34 & \phantom{0}9.33 & 11.20\\
Mean & 10.59 & 10.59 & 10.67 & 12.65 \\
3rd quartile & 12.08 & 12.24 & 12.15 & 14.70 \\
Max & 27.33 & 27.18 & 26.84 & 30.78
\\ \hline
\end{tabular*}
\end{table}

The APE measure results are not as conclusive as the RMSE. We see
that sometimes it is better to use half-hour intervals and sometimes
it is even better to use the hour intervals and not the 15-minutes
ones. However, it is also noticeable that the differences between
the three resolutions are usually quite small. The four-hour
resolution results are quite bad in comparison with the other
interval resolutions. These results can be used to justify the use
of half-hour intervals in our case study---only a minor improvement
to precision might be achieved by using a higher resolution.

In our comparisons, we have used the same $\operatorname{AR}(1)$ and $\operatorname{ARMA}(1,1)$
autocorrelation structure between periods no matter what their
length. (Of course, the parameters were estimated independently for
each.) A reviewer has pointed out, it is feasible that allowing a
different correlation structure for 15-minute periods may achieve
better performance than that based on half-hour periods. Due partly
to computational convergence problems for more complicated
correlation structures, these were not considered here.

\subsection{Dependence of precision on forecast
lead-time}\label{lead-time}

Our prediction process has three user defined elements: the learning
time; the prediction lead-time; and the forecasting horizon. During
our model's training stage we did not change these parameters.

Academic studies such as Weinberg, Brown and Stroud (\citeyear{JBayesian}) and Shen and Huang (\citeyear{SVDpredict})
concentrate on producing one-day-ahead predictions or sometimes
online updating forecasting algorithms. These methods, however, do
not address the industry problem of attaining good predictions in
order to produce the weekly schedule sufficiently ahead of time.
Trying to cope with this problem, our Telecom Company actually uses
a two stage process. It first produces a somewhat inaccurate (rough)
forecast ten days ahead of the desired week, and then it generates
another forecast five days before. The second forecast, we are being
told, is essential in order to adequately schedule agents. One
interesting question that arises from this practice is the extent to
which prediction lead-time effects forecasting accuracy.

In order to study prediction lead-time effects, we ran our
forecasting procedure using \textit{seven} different lead times,
ranging from one-day-ahead to seven-days-ahead. The learning period
and the forecasting horizon stay the same, that is, six weeks and one
day, respectively. We ran this prediction procedure using two
models: one is with a between-day covariance structure; and one
without it (just as we did in Section~\ref{random}). This allows us
to further examine the effect of this covariance structure.

We analyzed the prediction results of the two models for each
weekday separately. We do this in order to examine if certain
weekdays are more influenced by the change in forecast lead time.

\begin{figure}

\includegraphics{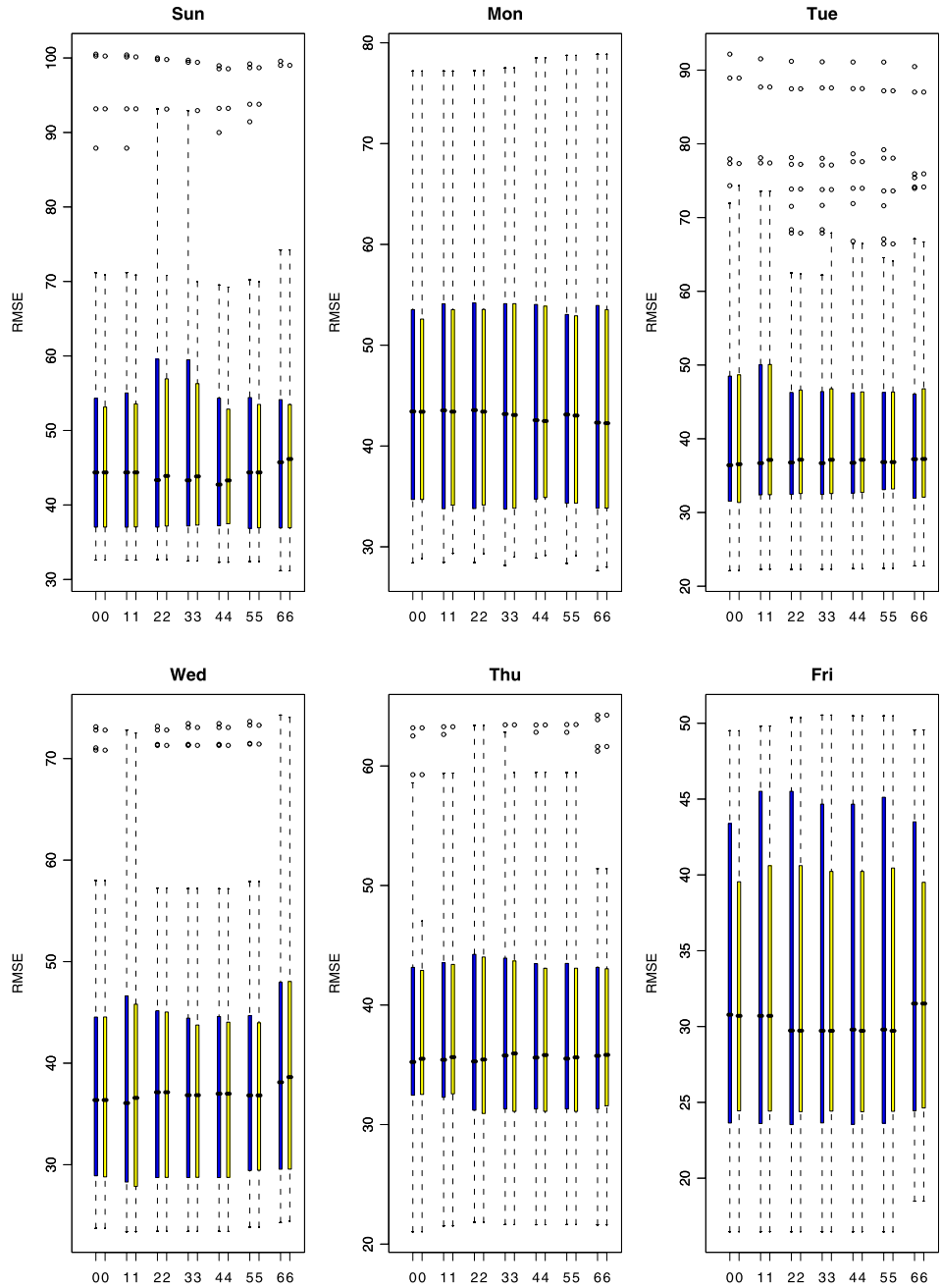}

\caption
{Comparison of RMSE as a function of the prediction lead time. The blue
(left) boxplot corresponds
to the model with $\operatorname{AR}(1)$ correlation structure between days and the
yellow (right) boxplot corresponds to the model without it.}\label{fig:rmsePredictionLag}
\end{figure}

\begin{figure}

\includegraphics{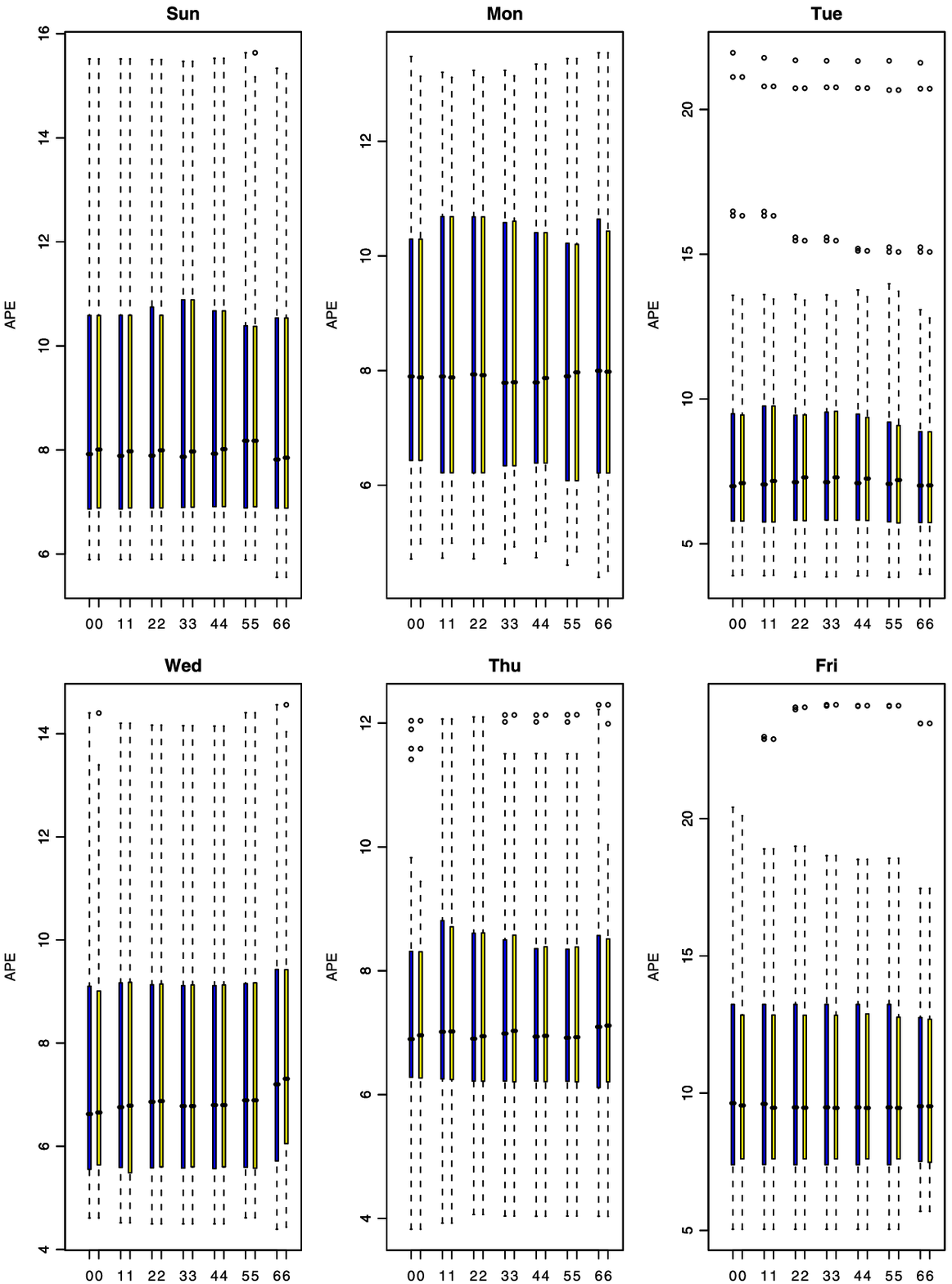}

\caption
{Comparison of APE as a function of the prediction lead time. The blue
(left) boxplot corresponds
to the model with $\operatorname{AR}(1)$ correlation structure between days and the
yellow (right) boxplot corresponds to the model without it.}\label{fig:apePredictionLag}
\end{figure}

Figures \ref{fig:rmsePredictionLag} and \ref{fig:apePredictionLag}
present the boxplots of individual RMSEs and APEs for each
weekday, as a function of the forecast lead time.

%

%

Surprisingly, the results indicate that shorter lead times do not
always improve predictions' accuracy. However, it is also important
to notice that the difference between the lead-time results for
each weekday are small for both the APE and the RMSE. This
property may be useful for call center managers who want to know
if they should update their week-ahead forecasts one or two days
ahead.

Another conclusion that can be derived is that incorporating the
$\operatorname{AR}(1)$ inter-day covariance structure generally improves the
prediction results. For example, looking at Figures~\ref{fig:rmsePredictionLag} and \ref{fig:apePredictionLag}, we can
see that the median RMSE and APE\footnote{The median is indicated
by a black horizontal short line inside the boxplots.} of the model
with the $\operatorname{AR}(1)$ covariance structure generally both have values less
than or equal to the median RMSE and APE of the model without it.

\section{Modeling expected service times}\label{section: service times}

In addition to predicting arrival rates, forecasting the workload of
a queuing system requires also predicting the average service
patterns (or alternatively the service rate pattern over each day).
Since our arrival's model applies a specific resolution (30 minutes),
one must predict average service times during those same
time intervals (periods).

Our model involves two explanatory variables that may affect service
rates: the weekday and the period. We compare two alternative
models where one is a generalization of the other.

In Figure \ref{fig:MeanServiceTimes}, the average service time
patterns resemble a quadratic curve.

Hence, the first model describes the average service time using a
quadratic regression in the periods, with interactions with the
weekday effect, where the period is included as a numeric variable
(rather than as a categorical variable). Intuitively speaking, this
model states that the daily service time curves differ among the
different weekdays but they are confined to be of a quadratic form.

This model formulates the average service time during period $k$,
that is, $z_{dk}$ in the following manner:
\begin{eqnarray}
\hspace*{7pt}\mbox{\textit{Model 1}:} \qquad z_{dk}&=&
\alpha_{q_{d}}+\beta_{1} \cdot k^2+ \beta_{2} \cdot k+\gamma_{1,q_{d}}\cdot k^2\nonumber
\\[-8pt]\\[-8pt]
&&{}+\gamma_{2,q_{d}}\cdot k+\phi\cdot{d}+\varepsilon_{dk};
\qquad\varepsilon_{dk} \sim N (0,\sigma^2),\nonumber
\end{eqnarray}
where $\alpha_{q}$ is the constant term related to the
$q$th weekday; $\beta_{1}$ and $\beta_{2}$ are,
respectively, the quadratic and linear coefficients;
$\gamma_{1,q_{d}}$ and $\gamma_{2,q_{d}}$ are the weekday-specific
quadratic and linear period effects and make up the weekday-period
interaction effects. The last effect is a postulated linear daily
trend coefficient denoted by $\phi$. We naturally added a random
error term denoted by $\varepsilon_{dk}$.

The second model is a generalization of the first model and it
assumes that the period's variable is a categorial variable. It
basically assumes that each weekday has its own average service
times pattern with no other restriction on its shape. Hence, it is a
generalization of the previous quadratic profile model. It is a more
complicated model which involves many more parameters. We add both
the linear daily trend effect to this model and the error term as
well. This model can be formulated in the following manner:
\begin{equation}
\mbox{\textit{Model 2}:} \qquad z_{dk}= \rho_{q_{d},k}+\phi\cdot{d}+\varepsilon_{dk};
\qquad \varepsilon_{dk} \sim N(0,\sigma^2),
\end{equation}
where $\rho_{q,k}$ is the interaction between the weekday and the
effect of the $k$th period.

We estimate parameters of the two average service times models using
the SAS\tsup{\textregistered} \textit{GLM} procedure. The learning data
include dates between mid-February, 2004 and the end of December,
2004. Examining the first model results shows that the interaction
between the quadratic period term and the weekday is not
significant. Consequently, we also examined the first model
excluding the insignificant term. This last model is referred to as
Model 3. Since our data has a large number of observations (6096
which correspond to 254 regular days), we use the asymptotic
log-likelihood ratio chi-square test to compare the models. Table~\ref{table:Service models results} summarizes the results of the
different models. We compare Model 3 against Model 2 to check if the
generalized model is significantly better than the reduced quadratic
model. The relevant chi-square statistic equals 13.524 and the
appropriate $p$-value is approximately one. It thus seems that the
generalized model (i.e., Model 2) is not significantly better in
modeling the average service times. Hence, we choose Model 3 as our
forecasting model.

\begin{figure}

\includegraphics{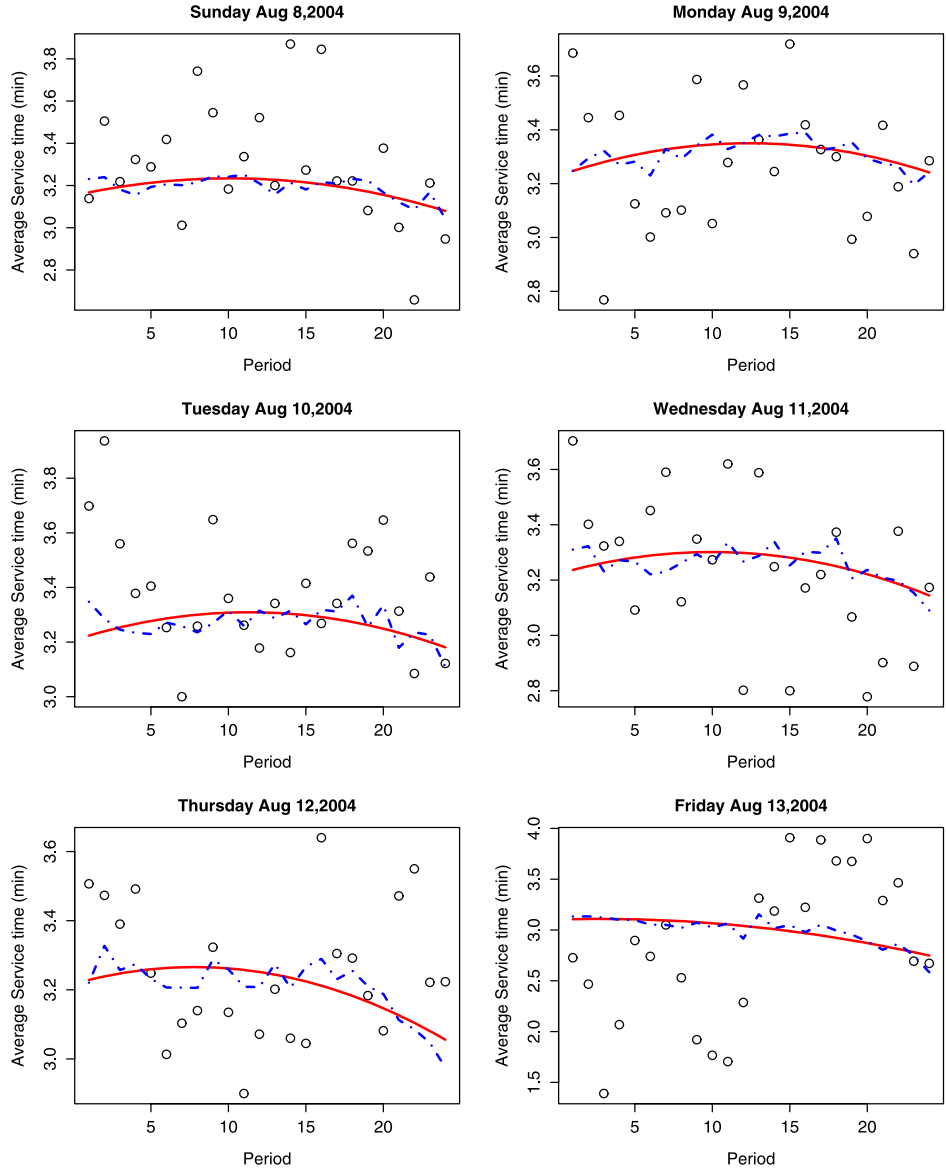}

\caption
{The predicted average service pattern for typical weekdays as a
function of period using Models \textup{2} and \textup{3}. The points
represent the (true) estimated average service times. The
(blue) dotted line represents the predicted average service times
using Model \textup{2} and the (red) line represents the appropriate predictions
using Model \textup{3}.}\label{fig:predService vs. period}
\end{figure}

Figure \ref{fig:predService vs. period} shows the predicted
pattern for each weekday between August 8, 2004 and August 13, 2004
versus the true average service times using Models 2 and 3.

By comparing the predictions to the true service means in the same
manner as we did with the arrival process analysis, we calculate
the mean APE. Its value is $7.68\%$. The predictions and the mean
APE value will later be used to estimate alternative measures of
system loads.

%
\begin{table}
\tablewidth=8.5cm
\caption
{Average service time models. Model \textup{1} assumes a different quadratic
curve for each weekday. Model \textup{3} is the same as Model \textup{1} excluding the
interaction between the quadratic period term and the weekday. Model
\textup{2} is the generalized model which assumes a different pattern for
each weekday}\label{table:Service models results}
\begin{tabular*}{\tablewidth}{@{\extracolsep{\fill}}lcc@{}}
\hline
\textbf{Model no.} & \textbf{No. of parameters} & \textbf{Error SS} \\
\hline
1 & \phantom{0}18 & 588.907 \\
2 & 144 & 575.884 \\
3 & \phantom{0}13 & 589.408 \\
\hline
\end{tabular*}
\end{table}

%

\section{Forecasting offered-loads in support of staffing}\label
{section:staffing}

In this section we begin with a brief review of how we obtained
estimators for the expected load using our data. We then introduce
measures that evaluate the performance of forecasts of the offered-load ($R$)
with respect to the QED ``square-root staffing'' rule. We also show
how to use these measures to evaluate the impact of prediction
errors (in the offered-loads) on call center operational
performance.

\subsection{Estimating expected load}

In an $M_t/G/N+G$ queueing system, the offered-load $R_t$ at time
$t$ is defined to be the average number of servers (agents) that
would be busy at time $t$, in the corresponding infinite server
system $M_t/G/\infty$ (with the same arrival process and same
service times). One can show [see, e.g., Whitt (\citeyear{queueing})]
that $R_t$ has the following two representations:
\begin{eqnarray}
R_t &=& E\int_{t-S}^{t} \lambda(u)\,du =
E\lambda(t-S_e)\cdot ES;
\end{eqnarray}
here $S$ is a generic service-time, and $S_e$ is a random variable
with the stationary-excess cdf associated with the service time cdf
G, namely,
\begin{eqnarray}
\operatorname{Pr}(S_e \leq t) \equiv\frac{1}{E(S)} \int_0^t [1-G(u)]\,du, \qquad t
\geq0.
\end{eqnarray}

The offered-load at time $t$ plays a central role in the planned
staffing at that time, as will be explained in the next section.
Whitt (\citeyear{queueing}) employs Taylor-series approximations to justify
the first-order approximation $R_t \approx\lambda(t-E(S_e))E(S)$.
If the system reaches a steady-state or if the arrival rate
$\lambda$ is constant, then the offered-load is exactly given by
$R=\lambda\cdot ES$. However, under time-varying arrivals, the two
representations above demonstrate that staffing at time $t$ must
take into account the arrivals prior to $t$ (over the stochastic
time interval $(t-S, t]$), which manifests itself through Whitt's
approximation [the arrival rate $\lambda$ at time $(t-E(S_e)$].

In our model we assume that the arrival process is an inhomogeneous
Poisson process but that during each 30-minute interval the arrival
rate remains constant. Hence, we predict the offered-load at day $d$
during the $k$th interval by the following natural
statistic:
\begin{equation} \tilde{R}_{dk}= \frac{\tilde{\lambda}_{dk} \cdot
\tilde{S}_{dk}}{30}, \qquad d=1,\ldots,D, k=1,\ldots,K,
\end{equation}
where $\tilde{\lambda}_{dk}$ is the predicted arrival count at day
$d$ during the $k$th interval, and $\hat{S}_{dk}$ is the
estimated average service time during that same interval. (The
division by 30 is to convert the predicted arrival rate into the
same units as the average service time---minutes). In concert with
the above, we also assume that performance during each time interval
can be predicted via the queueing model $M/G/N+G$.

Whitt (\citeyear{Whitt}) shows that, for practical parameter values, the
performance of the $M/G/N+G$ model is rather insensitive to the
service-time distribution, whereas the time to impatience
distribution has a far greater impact. This reduces performance
analysis to that of the tractable $M/M/N+G$ queue. But,
furthermore, in the desired regimes for call centers' operations,
$M/M/N+G$ can be in fact approximated (with a suitable
transformation of parameters) by $M/M/N+M$ (Erlang-A), as
articulated in Zeltyn (\citeyear{sergeyPhD}) and Manedelbaum and Zeltyn (\citeyear{MMGNconstraint}). These
observations will be used in the next section to evaluate system
performance.

\subsection{Error in $\beta$ and implications}\label{section:beta}

The square-root staffing rule, described in the \hyperref[sec1]{Introduction},
gives rise to the so-called Quality and Efficiency Driven (QED)
operational regime Gans Koole and Mandelbaum (\citeyear{CCtutorial}). Here, the number of
service providers (agents), denoted $N$, is determined by the
relation
\begin{equation}
N = \big\lceil R +\beta\cdot\sqrt{R} \big\rceil,
\end{equation}
where $R$ is the offered-load defined previously. The value of
$\beta$ determines a call center's operational regime: it is
Quality-Driven with large beta, Efficiency-Driven with small, and
QED if $\beta$ is near zero [typically within $(-2, 1)$]. In order
to maintain the latter regime, the system manager strives to achieve
a careful balance between service quality and efficiency. For a
discussion on how to determine $\beta$, as a function of operating
costs (staffing and congestion costs), see~Manedelbaum and Zeltyn (\citeyear{MMGNconstraint}), Zeltyn (\citeyear{sergeyPhD}), Borst, Mandelbaum and Reiman (\citeyear{dimensioning}).

Define $\beta_u$ as the \textit{user} (say, call center manager)
chosen $\beta$. In order to forecast the number of required
agents, $\tilde{N}$, the user will use the predicted values of the
arrival and service rates to forecast the offered-load, and set
\begin{equation}\label{user_beta}
\tilde{N}=\tilde{R}+\beta_u \cdot\sqrt{\tilde{R}}, \\
\end{equation}
where $\tilde{R}$ is the forecasted offered-load.

In practice, the assigned agents will face the true (realized)
value ($\lambda$) of the arrival rate and the actual service rate
($\mu$). With these real values of $\lambda$, $\mu$,\vspace*{1pt} and the
corresponding value of $R$ and the pre-determined number of agents
$\tilde{N}$, the call center is in effect operating under a
different value of $\beta$. This \textit{adjusted} value of $\beta$
will be referred to as $\widetilde{\beta_{a}}$, which is determined
via
\[
\widetilde{\beta_{a}} = \frac{\tilde{N}-R}{\sqrt{R}}.
\]

The above expression can be rewritten as
\begin{equation}\label{adjusted_beta}
\tilde{N} = R+\widetilde{\beta_{a}}\sqrt{R}.
\end{equation}

Since the number of pre-assigned agents ($\tilde{N}$) is the same in
both equations (\ref{user_beta}) and (\ref{adjusted_beta}), we can
equate them and conclude the following:
\begin{eqnarray}
\nonumber R +\widetilde{\beta_{a}} \cdot\sqrt{R}
&=&\tilde{R}+\beta_u \cdot\sqrt{\tilde{R}}, \\
\nonumber\widetilde{\beta_a}-\beta_u \cdot
\sqrt{\frac{\tilde{R}}{R}}&=& \frac{\tilde{R}-R}{\sqrt{R}}.
\end{eqnarray}

The mean APE for average service time is 0.0768, which means that
$\sqrt{\frac{\tilde{\mu}}{\mu}}$ has a value of approximately 1. The
mean APE\vspace*{2pt} arrival rate is 0.0927, indicating that
$\sqrt{\frac{\tilde{\lambda}}{\lambda}}$ also has a value of
approximately 1. Considering these two values, we conclude that
$\sqrt{\frac{\tilde{R}}{R}} \approx1$, which allows us to make the
following approximation:
\begin{equation}
\Delta\beta\triangleq\widetilde{\beta_{a}}-\beta_u {\approx}\
\frac{\tilde{R}-R}{\sqrt{R}}.
\end{equation}

The quantity $\Delta\beta$ measures the standardized difference
between the predicted offered-load and the true offered-load.
Examining its values can help assess forecasting quality. It enables
one to answer questions such as follows: does this forecasting algorithm
usually overestimate or under-estimate the number of arrivals?;
and, by how many agents will one under or overstaff? Note that the
ideal value of $\Delta\beta$ is zero, indicating a perfect point
prediction of the realized offered-load.

\begin{figure}

\includegraphics{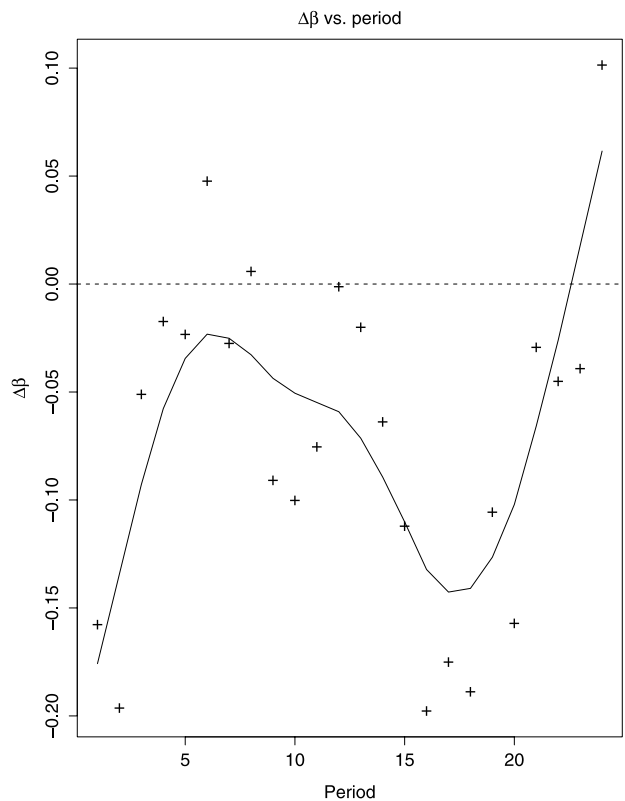}

\caption
{The average estimated $\Delta\beta$ as a function of period.}\label{fig:Beta vs. period}
\end{figure}

In Figure \ref{fig:Beta vs. period} we examine the averaged $\Delta
\beta$ values across the 24 periods of the day using our final mixed
model predictions. To obtain these averaged values of the estimated
$\Delta\beta$, we first estimate $\Delta\beta$ for each day in our
learning data during each period. These were averaged for each
period separately over all the days, excluding holidays and
irregular days.

%

Small absolute values of $\Delta\beta$ (a user's $\beta_{u}$ close
to the actual $\widetilde{\beta_{a}}$) indicate that our model does
quite well in predicting the values of offered-load. The estimated
average values are close to zero but are usually smaller than zero,
which means that for most parts of the day the predictions would
lead to some under-staffing.

\begin{figure}

\includegraphics{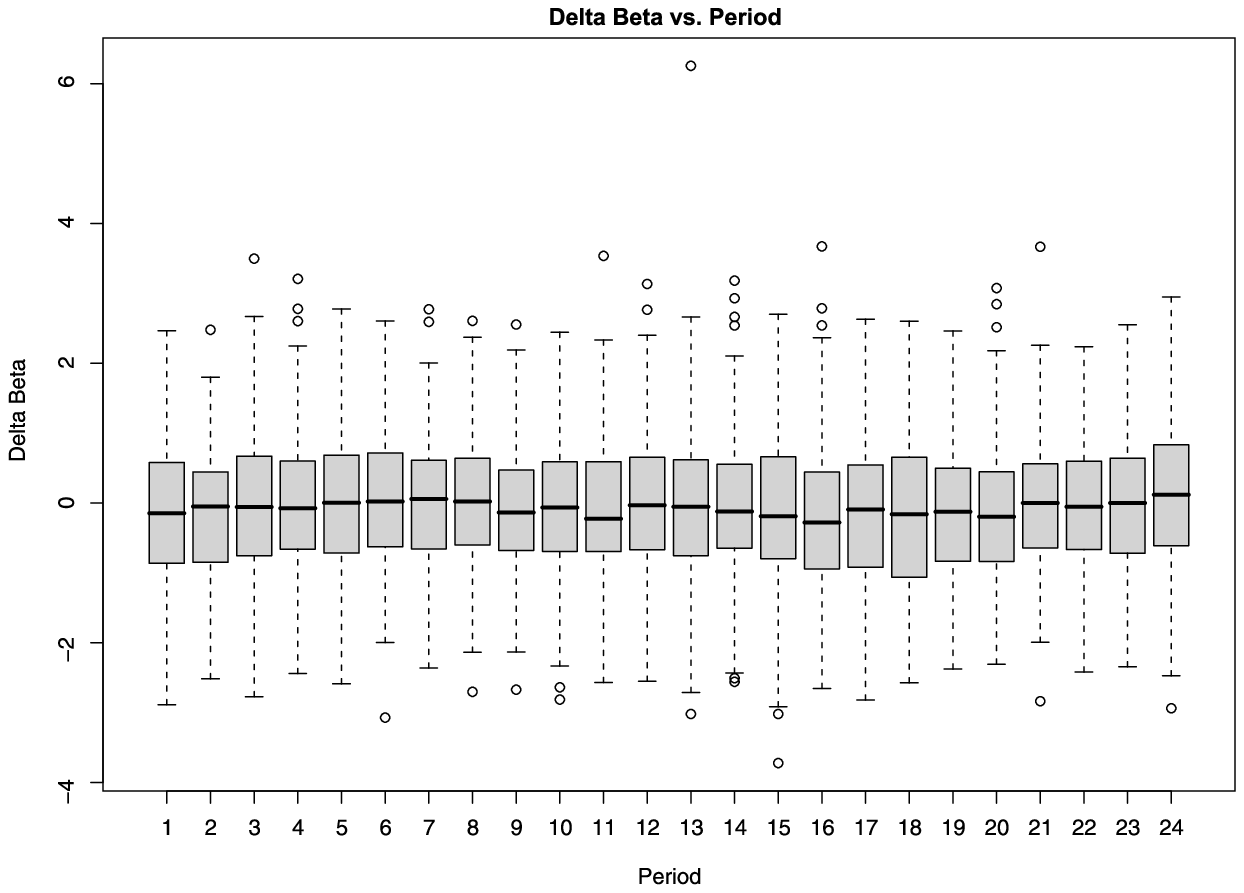}

\caption
{Boxplots of $\Delta\beta$ for the different periods.}\label{fig:Beta boxplot}
\end{figure}

After examining the values of average $\Delta\beta$, one can also
examine the dispersion of $\Delta\beta$ throughout the day. To
this end, consider the boxplots for each period, as shown in
Figure \ref{fig:Beta boxplot}. We note that most of the $\Delta
\beta$ values have absolute value less than 1. (There seems to be
an outlier in the $13$th period. This corresponds to one
day, October $10$th, where the average service time was
particularly short for some unexplained reason.) The boxplots
reveal that $50\%$ of $\Delta\beta$ values are between $-$0.75 and
0.75 and on average are zero. Given that practical values of
$\beta$ are between $-$1 and 2, on a large fraction of the periods
our mixed model will not make a gross error in predicting the
offered-load.

%

We now proceed to analyze how deviations in $\Delta\beta$ affect a
call center's operational performance. Specifically, we examine
three performance measures: the probability of being delayed for
service, $\operatorname{Pr}(W>0)$; the probability of abandonment, $\operatorname{Pr}(Ab)$; and
the average waiting time, $E(W)$. We then compute these measures in
an $M/M/N+G$ environment, using the asymptotic results in
Zeltyn (\citeyear{sergeyPhD}), Manedelbaum and Zeltyn (\citeyear{MMGNconstraint}). The relevant formulae require
knowledge of the value of $\beta$, as well as the ratio between the
service rate $\mu=1/E(S)$ and an (im)patience rate parameter
$\theta$. [One can think of $1/\theta$ as some measure of average
(im)patience.]

We compute performance measures for three values of the ratio
between service rate and patience rate: 0.1---corresponding to
impatient customers; 1---average customer patience is the same as
their average service time; and 2---implying that customers are
rather patient. (Based on our experience, a value of 2 for this
ratio is quite realistic.) For further discussion on these issues
refer to~Zeltyn (\citeyear{sergeyPhD}), Manedelbaum and Zeltyn (\citeyear{MMGNconstraint}).

Since $\Delta\beta$ only provides us with the deviation between
$\beta_a$ and $\beta_u$, we must determine one of them in order to
compute the other. Hence, we choose three reasonable values for
$\beta_u$: $-1,0,1$. Using these values and $\Delta\beta$, we can
calculate $\beta_a$. The next step is to compute performance
measures using the two different $\beta$'s, the predicted
offered-load and the observed offered-load. This provides us with
the three performance measures for each period during each day, for
$\beta_u$ and $\beta_a$.

To analyze the outcomes, we averaged each performance measure over
the 203 days. This provides us with all three average performance
measures for each period using $\beta_u$ and $\beta_a$, and for each
of the three different settings of the ratio between $\mu$ and
$\theta$. Tables \ref{Table: Performance Measures} and \ref{Table:
Performance Measures2} show the average performance measures for two
selected periods: the 6th, 13:00--13:30, and the
16th, 18:00--18:30. (When looking at Figure~\ref{fig:Beta vs. period},
one observes that the smallest average
absolute $\Delta\beta$ occurs around the 6th~period and
the largest deviation occurs at around the 16th, which
is why these two periods were chosen.)

%
\tabcolsep=0pt
\begin{table}
\caption
{Average Performance Measures for the period between 13:00--13:30
using $\beta_u=-1,0,1$ and $\mu/ \theta=0.1,1,2$}\label{Table:
Performance Measures}
\begin{tabular*}{\textwidth}{@{\extracolsep{\fill}}l d{1.3}d{1.3}d{1.3} d{1.3}d{1.3}d{1.3} d{1.4}d{2.3}d{2.3}@{}}
\hline
&\multicolumn{3}{c}{$\bolds{\operatorname{Pr}(W>0)}$}
&\multicolumn{3}{c}{$\bolds{\operatorname{Pr}}$\textbf{(Abandon)}}
& \multicolumn{3}{c@{}}{$\bolds{E(W)}$ \textbf{(sec)}} \\[-6pt]
&\multicolumn{3}{c}{\hrulefill}
&\multicolumn{3}{c}{\hrulefill}
& \multicolumn{3}{c@{}}{\hrulefill} \\
& \multicolumn{1}{c}{$\bolds{\frac{\mu}{\theta}=0.1}$}
& \multicolumn{1}{c}{$\bolds{\frac{\mu}{\theta}=1}$}
& \multicolumn{1}{c}{$\bolds{\frac{\mu}{\theta}=2}$}
& \multicolumn{1}{c}{$\bolds{\frac{\mu}{\theta}=0.1}$}
& \multicolumn{1}{c}{$\bolds{\frac{\mu}{\theta}=1}$}
& \multicolumn{1}{c}{$\bolds{\frac{\mu}{\theta}=2}$}
& \multicolumn{1}{c}{$\bolds{\frac{\mu}{\theta}=0.1}$}
& \multicolumn{1}{c}{$\bolds{\frac{\mu}{\theta}=1}$}
& \multicolumn{1}{c@{}}{$\bolds{\frac{\mu}{\theta}=2}$}\\
\hline
$\beta_u=-1$ & 0.442 &0.841 & 0.931 & 0.199 & 0.167 &0.160 & 3.882& 32.528 & 62.223
\\[1pt]
$\widetilde{\beta_a}$ & 0.428 & 0.754 & 0.819 & 0.210 & 0.180 & 0.174 & 4.099 & 35.741 & 69.173 \\
Ratio & 0.968 & 0.896 & 0.900 & 1.055 &1.077& 1.087& 1.055& 1.098 & 1.111
\\[3pt]
$\beta_u=0$ & 0.240 & 0.5 & 0.586 & 0.087 & 0.057 & 0.047 & 1.7023 &11.091 & 18.377
\\[1pt]
$\widetilde{\beta_a}$ & 0.252 & 0.487 & 0.550 & 0.010& 0.076& 0.070 &1.993 & 15.270 & 27.94\\
Ratio & 1.05 & 0.974 & 0.938 & 0.115 & 1.333 & 1.490 &1.171 & 1.377& 1.520
\\[3pt]
$\beta_u=1$ & 0.083 & 0.159 & 0.179 & 0.024 & 0.011& 0.008 & 0.476 &2.173 & 2.975
\\[1pt]
$\widetilde{\beta_a}$ & 0.113 & 0.224 & 0.257 & 0.038 & 0.024 & 0.020& 0.760 & 4.850 & 8.558\\
Ratio & 1.361 & 1.409 & 1.436 & 1.583& 2.182 & 2.500 & 1.596 &2.232 & 2.876 \\
\hline
\end{tabular*}
\vspace*{5pt}
\end{table}

%
\tabcolsep=0pt
\begin{table}[b]
\caption
{Average Performance Measures for the period between 18:00--18:30
$\beta_u=-1,0,1$ and $\mu/ \theta=0.1,1,2$}\label{Table:
Performance Measures2}
\begin{tabular*}{\textwidth}{@{\extracolsep{\fill}}ld{1.3}d{1.3}d{1.3}d{1.3}d{1.3}d{1.4}d{1.3}d{2.3}d{2.3}@{}}
\hline
&\multicolumn{3}{c}{$\bolds{\operatorname{Pr}(W>0)}$}
&\multicolumn{3}{c}{$\bolds{\operatorname{Pr}}$\textbf{(Abandon)}}
& \multicolumn{3}{c@{}}{$\bolds{E(W)}$ \textbf{(sec)}} \\[-6pt]
&\multicolumn{3}{c}{\hrulefill}
&\multicolumn{3}{c}{\hrulefill}
& \multicolumn{3}{c@{}}{\hrulefill} \\
& \multicolumn{1}{c}{$\bolds{\frac{\mu}{\theta}=0.1}$}
& \multicolumn{1}{c}{$\bolds{\frac{\mu}{\theta}=1}$}
& \multicolumn{1}{c}{$\bolds{\frac{\mu}{\theta}=2}$}
& \multicolumn{1}{c}{$\bolds{\frac{\mu}{\theta}=0.1}$}
& \multicolumn{1}{c}{$\bolds{\frac{\mu}{\theta}=1}$}
& \multicolumn{1}{c}{$\bolds{\frac{\mu}{\theta}=2}$}
& \multicolumn{1}{c}{$\bolds{\frac{\mu}{\theta}=0.1}$}
& \multicolumn{1}{c}{$\bolds{\frac{\mu}{\theta}=1}$}
& \multicolumn{1}{c@{}}{$\bolds{\frac{\mu}{\theta}=2}$} \\
\hline
$\beta_u=-1$ & 0.442 & 0.841 & 0.931 & 0.214 & 0.179 & 0.172 & 4.100&34.355 &65.716
\\[1pt]
$\widetilde{\beta_a}$ & 0.475 & 0.798 & 0.854 & 0.172 & 0.230 & 0.225& 5.182 & 46.597 & 91.144 \\
Ratio & 1.05 & 0.989 & 0.917 & 0.804 & 1.285 & 1.308 & 1.264 & 1.356 &1.387
\\[3pt]
$\beta_u=0$ & 0.240 & 0.5 & 0.586 & 0.092 & 0.060 & 0.0501 & 1.764 &11.606 & 19.229
\\[1pt]
$\widetilde{\beta_a}$ & 0.299 & 0.562 & 0.627 & 0.131 & 0.105 & 0.098& 2.666 & 21.556 & 40.434 \\
Ratio & 1.246 & 1.124 & 1.070 & 1.424 & 1.750 & 1.956 & 1.511 & 1.857 & 2.103
\\[3pt]
$\beta_u=1$ & 0.083 & 0.159 & 0.179 & 0.025 & 0.012& 0.008 & 0.490 &2.255 & 3.095
\\[1pt]
$\widetilde{\beta_a}$ & 0.149 & 0.295 & 0.337 & 0.054 & 0.037 & 0.032& 1.116 & 7.699 & 13.395\\
Ratio & 1.795 & 1.855 & 1.883 & 2.160 & 3.083 & 4.000 & 2.278 &3.414& 4.328 \\
\hline
\end{tabular*}
\vspace*{5pt}
\end{table}

%
\begin{figure}

\includegraphics{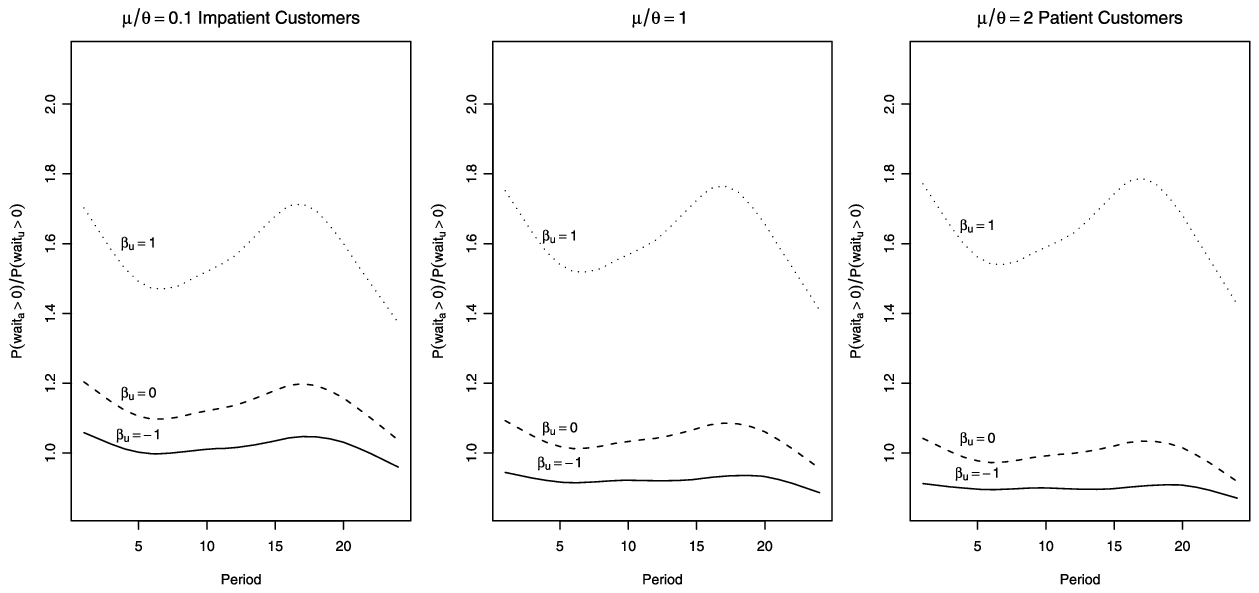}

\caption
{The average waiting probability ratio over periods. Each of the three
plots has
three curves. Each curve corresponds to a different $\beta_u$ value
(i.e., $-1, 0, 1$). The
plots examine the ratio of the average waiting probabilities using
$\widetilde{\beta_a}$ and $\beta_u$. The three plots, from left to
right, show
the results when using ratios of $\frac{\mu}{\theta}=0.1$, $\frac
{\mu}{\theta}=1$ and $\frac{\mu}{\theta}=2$, respectively.}\label{fig:waitp}
\end{figure}

%
\begin{figure}[b]

\includegraphics{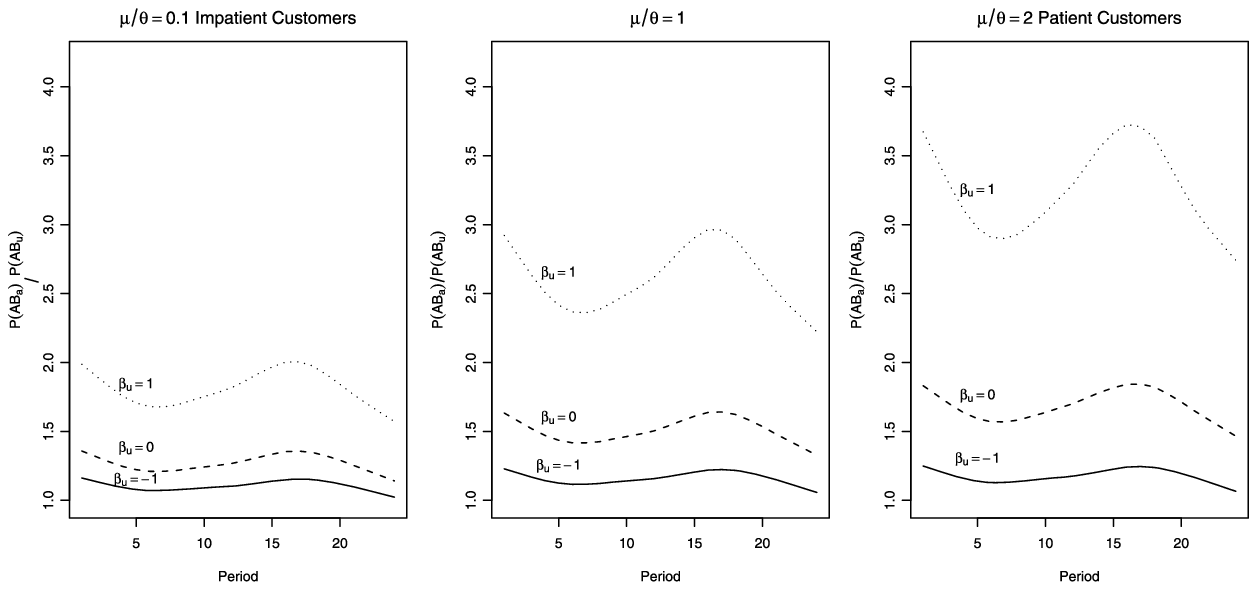}

\caption
{The average abandonment probability ratio over periods. Each of the
three plots has
three curves.\vspace*{1pt} Each curve corresponds to a different $\beta_u$ value
(i.e., $-1, 0, 1$). The
plots examine the ratio of the average abandonment probabilities using
$\widetilde{\beta_a}$ and $\beta_u$. The three plots, from left to
right, show
the results when using ratios of $\frac{\mu}{\theta}=0.1$, $\frac
{\mu}{\theta}=1$ and $\frac{\mu}{\theta}=2$, respectively.}\label{fig:Pab}
\end{figure}

\begin{figure}

\includegraphics{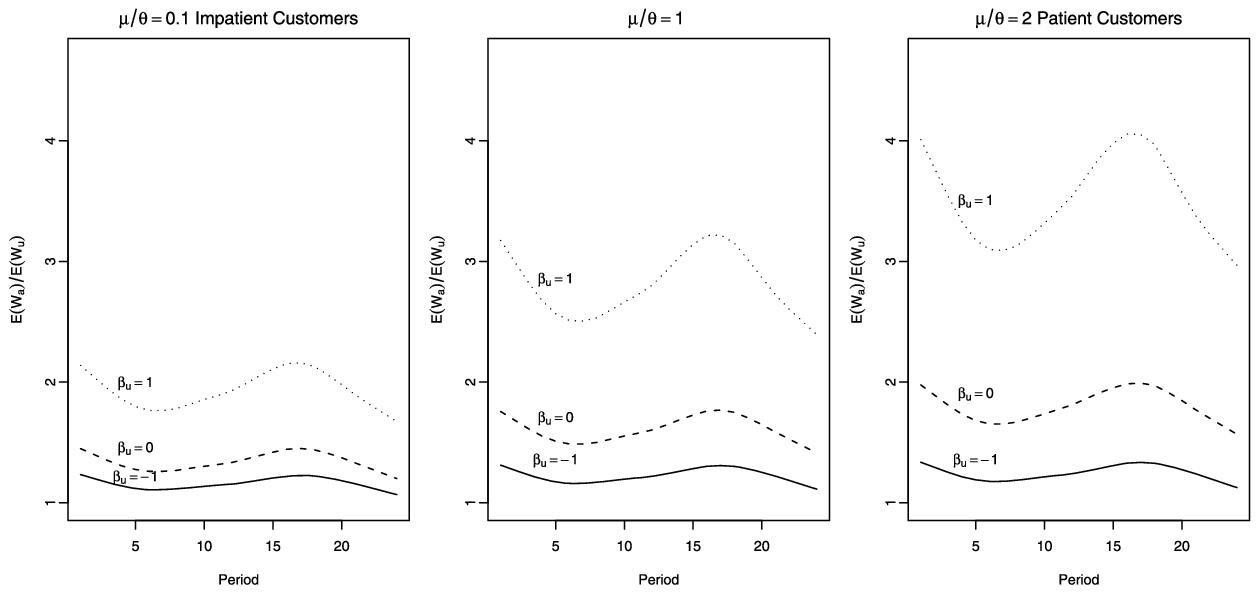}

\caption
{The average waiting time ratio over periods. Each of the three plots has
three curves. Each curve corresponds to a different $\beta_u$ value
(i.e., $-1, 0, 1$). The
plots examine the ratio of the average waiting time using $\widetilde
{\beta_a}$ and $\beta_u$. The three plots, from left to right, show
the results when using ratios of $\frac{\mu}{\theta}=0.1$, $\frac
{\mu}{\theta}=1$ and $\frac{\mu}{\theta}=2$, respectively.}\label{fig:Ew}
\end{figure}

 In Figures~\ref{fig:waitp}, \ref{fig:Pab} and
\ref{fig:Ew}, we summarize the ratios between the three performance
measures, using $\widetilde{\beta_a}$ and $\beta_u$. It is clear
that the ratios of each measure vary along the day. The graphs
examine not only how the ratios change when the user (call center manager)
chooses different values of $\beta_u$ but also when the ratio
between the service rate, $\mu$, and the (im)patience rate,
$\theta$, varies. The following conclusions can be drawn from the
graphs and tables:
\begin{itemize}
\item The average waiting probability ratios show more variability for
higher values of
$\beta_u$. However, when we examine the average waiting
probability ratios across different ratios of $\mu$ and
$\theta$, there is little variation.
\item The average abandonment probability ratios vary more for higher
values of
$\beta_u$, and also for higher values of the ratio between
$\mu$ and $\theta$. From the tables we see that the
probabilities of abandonment are not large (usually smaller
than 0.1).
\item The average waiting time ratios vary more for higher values of
$\beta_u$ and this is also true for higher values of the
ratio between $\mu$ and $\theta$. From the tables we see
that the average waiting times are not long (usually less
than 1 minute).
\end{itemize}

%

\subsection{Error in number of servers}

The actual error in the number of servers, due to forecasting
error, is also of interest. Using the same methodology as before,
one can assume that $\beta$ has been chosen to achieve specified
quality and efficiency goals. Then, based on the predicted values
$\tilde{R}$ of the true offered-load $R$, the number of required
agents $\tilde{N}$ is estimated via
\begin{equation}
\tilde{N}= \tilde{R}+\beta\cdot\sqrt{\tilde{R}}.
\end{equation}

Knowing the true offered-load, $R$, the appropriate number of
agents is $N$---the correct number of agents required to handle
the actual system load at the desired quality and efficiency
trade-off. Formally,
\begin{equation}
N= R+\beta\cdot\sqrt{R}.
\end{equation}
From the above two formulae and the conclusions that were derived in
Section~\ref{section:beta}, one deduces the following relations:
\begin{eqnarray}\label{Equation:service providers delta} \Delta N \triangleq
{\hat{N}-N} \approx\tilde{R}-R \approx\Delta\beta\cdot\sqrt{R}
.
\end{eqnarray}

The measure $\Delta N$ enables one to evaluate the difference
between the actually deployed and the desired number of agents.

\begin{figure}

\includegraphics{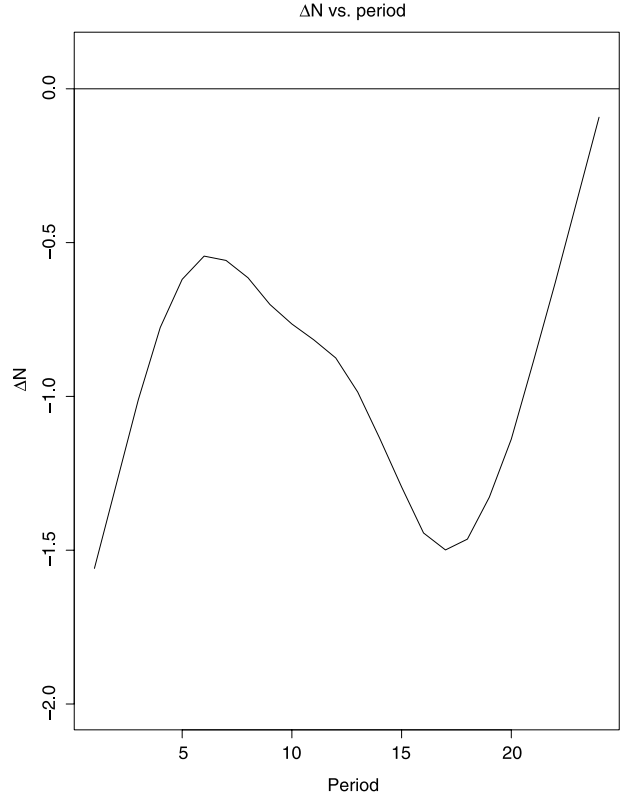}

\caption
{The average estimated $\Delta N$ as a function of period.}\label{fig: Delta A vs. period}
\end{figure}

We proceed with calculating $\Delta N$, using the same data as
before. Figure \ref{fig: Delta A vs. period} demonstrates the
results.

%

Not surprisingly, the outcomes are quite similar to the graphs we
observed when examining $\Delta\beta$. We are underestimating the
offered-load and hence under-staffing the call center. However, the
deviation between the required and the predicted number of agents
for most periods is less than 1 agent, which is encouraging. This
seems to indicate that our prediction models offer a rather good
solution for estimating the offered-load in the QED regime.

\subsection{Lead-time effect}

In Section~\ref{lead-time} we examined the effect of forecast
lead time on the precision of arrival counts predictions. The
analysis revealed that the forecast lead time has no significant
influence on precision. Now we would like to examine how the
forecast lead time affects $\Delta\beta$ and $\Delta N$.
Intuitively, one would expect both of these measures to decrease in
their absolute value as we shorten lead times.

\begin{figure}

\includegraphics{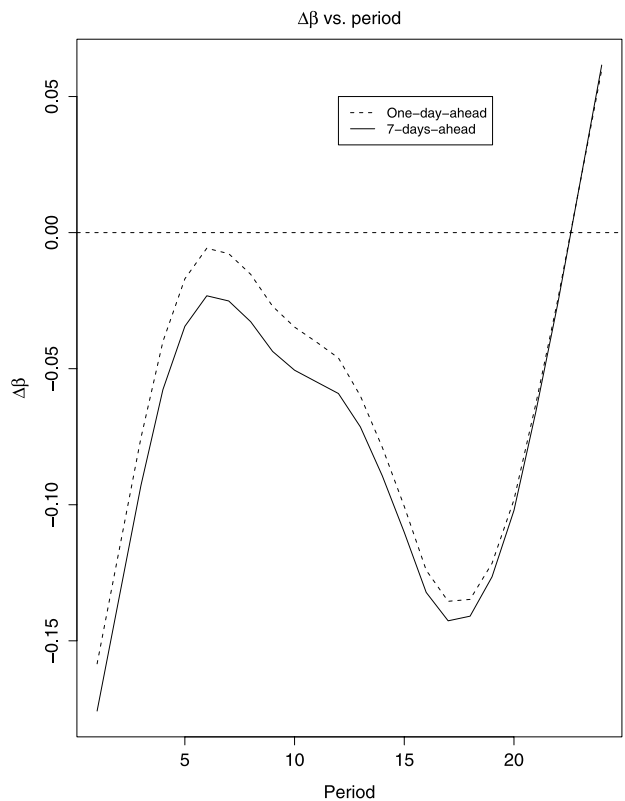}

\caption
{The average estimated $\Delta\beta$ as a function of period for
different forecast lead times.}\label{fig:LT beta}
\end{figure}

\begin{figure}

\includegraphics{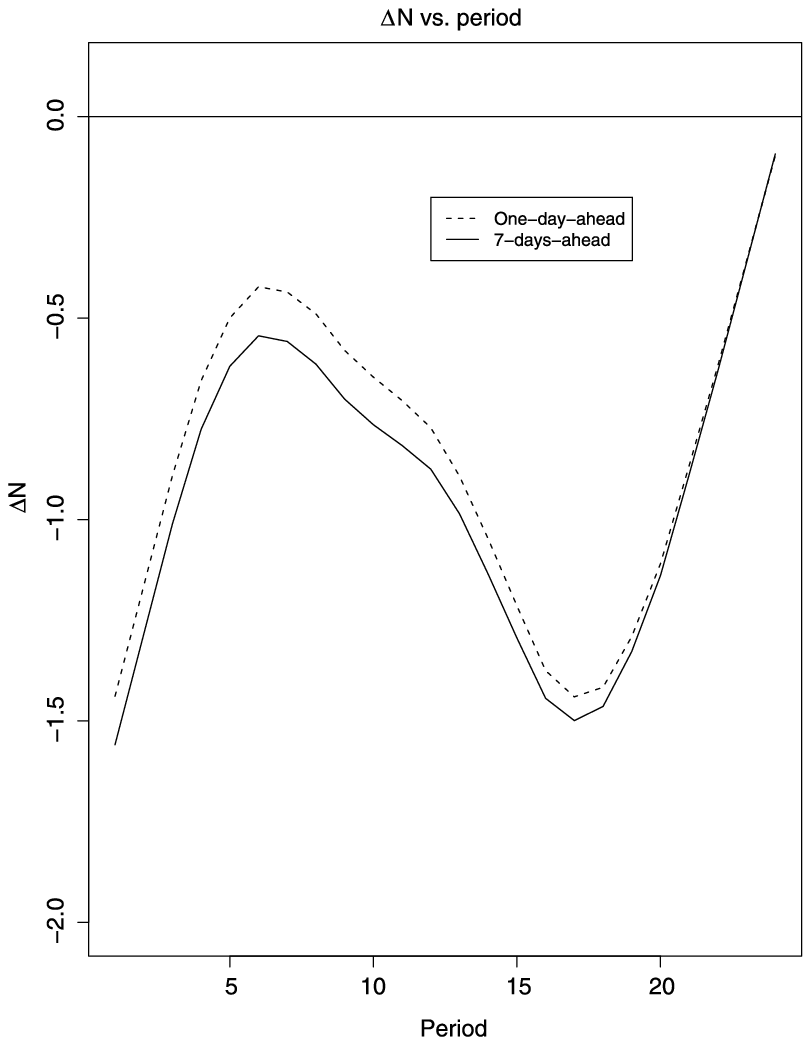}

\caption
{The average estimated $\Delta N$ as a function of period for different
forecast lead times.}\label{fig: LT Delta S}
\end{figure}

Figure \ref{fig:LT beta} shows the smoothed average $\Delta\beta$
for each period, using one-day-ahead predictions and a week-ahead
predictions. Figure \ref{fig: LT Delta S} presents the average
$\Delta N$ under the two different forecast lead-time alternatives.

The differences between the average results of the two measures are
not very significant. However, we observe that the one-day-ahead
results are better than the week-ahead results since the absolute
values of the deviations in both $\Delta\beta$ and $\Delta N$ are
smaller. Also, both measures indicated that the week-ahead
predictions were fairly accurate to begin with and, hence, we did not
expect them to improve by much.

%

%

\section{Conclusions and further research}

Our mixed model was customized to the specific requirements of an
Israeli Telecom company. The company requires that the weekly
forecast be available to the decision makers a few days in advance
and should be based on six weeks of past data. Recent research, on
the other hand, has focused on producing one-day-ahead forecasts or
within-day learning algorithms. These issues are very important and
may be very useful for call centers that can mobilize their agents
on short notice. As we show, our mixed model does contain the much
needed practical flexibility to also support long lead times and
short learning periods. It is also relatively easy to implement with
standard software such as SAS\tsup{\textregistered} and yields a
computationally efficient solution.

The mixed model incorporates fixed effects, such as day-of-week and
its interaction with the daily periods; but it also models the
day-to-day and the period-to-period correlations. We have detailed
how to determine the significance of such effects. The modeling
approach allowed us to incorporate billing cycle information as
exogenous variables after a preliminary examination was carried out
to help reduce the parameter vector dimension.

Unfortunately, we could not obtain the Israeli Telecom company's
predictions for the learning data used in this paper since the
company does not regularly maintain its past predictions. However,
we were able to attain the company's predictions for the week
between November 4th, 2007 and November
9th, 2007. Based on these predictions, it seems that
our mixed model and the company's algorithm both produce similar
results for most periods of the day. Yet, during periods of the
day which exhibit more instability (mainly early morning and late
evening), their algorithm performs worse than our mixed model.

The mixed model results were examined under different lead times.
From a practical perspective, a manager might wish to consider a
two-stage prediction forecast: first, producing an early weekly
forecast for the scheduling process; and next, producing another
forecast one day before the week begins. This later prediction might
provide a more reliable forecast. Using this one-day-ahead forecast,
the manager of a call center might be able to incorporate immediate
changes into the weekly schedule.

We investigated the interval resolution effect on the prediction
accuracy. This analysis showed that, for practical implementations, one
can use a half-hour resolution with only little loss to the
prediction accuracy levels. Using this resolution can simplify the
problem when compared with the 15-minute resolution. It is also a
practical time interval which call center managers often use for
determining their weekly staff schedule.

Our original forecast problem was to predict the system load based
on the average service times and the arrival rates. To model the
average service times, we suggested and fitted a fairly easy
quadratic regression model which incorporates weekdays and period
effects.

Having both the average service times' predictions and the arrival
counts, we constructed QED regime performance measures. Using these
measures, we analyzed the effect of our models results on a
call center's operational performance (i.e., the probability of
being delayed for service, the probability of abandonment and the
average waiting time). Our results indicate that during busy
periods, when the QED regime's ``square-root staffing'' rule is
relevant, the system will perform at a desired level commensurate
with the actual load or very close to it. This result gives some
evidence that one can ensure a prespecified QED regime using load
forecasts that are sufficiently precise.

An interesting future research question is how to estimate and
predict the system load function $R(t)$ directly (instead of
predicting the average service times and arrival rates separately).
Such estimation problems can be approached using nonparametric
methods or parametric models. It will be interesting to learn if
such prediction methods yield better results than the method
proposed here. [For further discussion about the effects of
time-varying system load on determining staffing levels, the reader
is referred to Feldman et~al. (\citeyear{Zohar}).]

\section*{Acknowledgments}
The authors thank Professor Larry Brown for proofreading and helpful
suggestions. We also want to thank a number of reviewers for
constructive comments and suggestions.

\begin{supplement}[id=suppA]
\sname{Supplement}
\stitle{Israeli Telecom company call center data set and forecasting code}
\slink[doi]{10.1214/09-AOAS255SUPP}
\slink[url]{http://lib.stat.cmu.edu/aoas/255/Supplement.zip}
\sdatatype{.zip}
\sdescription{The zip folder contains three files: a readme file which
describes the data set in detail; the Israeli Telecom company data set
both in text format and SAS format; and the SAS\tsup{\textregistered}/STAT
[SAS (\citeyear{SAS})] code which was used to create our final forecasting model.}
\end{supplement}

\printaddresses

\end{document}